\documentclass[acmsmall]{acmart}

\usepackage{amsmath,amsfonts}
\usepackage{algorithmic}
\usepackage{graphicx}
\usepackage{textcomp}
\usepackage{xcolor}
\usepackage{multirow}
\usepackage{caption}
\usepackage{subcaption}
\usepackage{siunitx}

\definecolor{darkgreen}{RGB}{0,0,0}
\newcommand{\blue}[1]{\textcolor{darkgreen}{#1}}

\newcommand{\green}[1]{{\color{black}#1}}

\usepackage[linesnumbered,ruled,vlined]{algorithm2e}

\AtBeginDocument{%
  \providecommand\BibTeX{{%
    \normalfont B\kern-0.5em{\scshape i\kern-0.25em b}\kern-0.8em\TeX}}}

\setcopyright{rightsretained}
\acmJournal{TECS}
\acmYear{2021} \acmVolume{1} \acmNumber{1} \acmArticle{1} \acmMonth{1} \acmPrice{}\acmDOI{10.1145/3477006}

\acmJournal{TECS}
\acmVolume{0}
\acmNumber{0}
\acmArticle{0}
\acmMonth{10}

\begin{document}

\title{SAGE: A \underline{S}plit-\underline{A}rchitecture Methodolo\underline{g}y for \underline{E}fficient End-to-End Autonomous Vehicle Control}


\author{Arnav~Malawade}
\authornote{Both authors contributed equally to this research.}
\email{malawada@uci.edu}
\orcid{0000-0002-4974-1622}
\affiliation{%
  \institution{University of California Irvine}
  \city{Irvine}
  \country{USA}
}
\author{Mohanad~Odema}
\authornotemark[1]
\email{modema@uci.edu}
\orcid{0000-0002-0828-949X}
\affiliation{%
  \institution{University of California Irvine}
  \city{Irvine}
  \country{USA}
}
\author{Sebastien~Lajeunesse-DeGroot}
\email{slajeune@uci.edu}
\orcid{0000-0002-9966-8435}
\affiliation{%
  \institution{University of California Irvine}
  \city{Irvine}
  \country{USA}
}
\author{Mohammad~Abdullah~Al~Faruque}
\email{alfaruqu@uci.edu}
\orcid{0000-0002-5390-0497}
\affiliation{%
  \institution{University of California Irvine}
  \city{Irvine}
  \country{USA}
}

\thanks{This article appears as part of the ESWEEK-TECS special issue and was presented in the International Conference on Hardware/Software Codesign and System Synthesis (CODES+ISSS), 2021.}


\begin{abstract}
Autonomous vehicles (AV) are expected to revolutionize transportation and improve road safety significantly.
However, these benefits do not come without cost; AVs require large Deep-Learning (DL) models and powerful hardware platforms to operate reliably in real-time, requiring between several hundred watts to one kilowatt of power. This power consumption can dramatically reduce vehicles' driving range and affect emissions. 
To address this problem, we propose SAGE: a methodology for selectively offloading the key energy-consuming modules of DL architectures to the cloud to optimize edge energy usage while meeting real-time latency constraints. 
Furthermore, we leverage Head Network Distillation (HND) to introduce efficient \textit{bottlenecks} within the DL architecture in order to minimize the network overhead costs of offloading with almost no degradation in the model's performance. 
We evaluate SAGE using an Nvidia Jetson TX2 and an industry-standard Nvidia Drive PX2 as the AV edge devices and demonstrate that our offloading strategy is practical for a wide range of DL models and internet connection bandwidths on 3G, 4G LTE, and WiFi technologies. Compared to edge-only computation, SAGE reduces energy consumption by an average of \textbf{36.13\%}, \textbf{47.07\%}, and \textbf{55.66\%} for an AV with one low-resolution camera, one high-resolution camera, and three high-resolution cameras, respectively. SAGE also reduces upload data size by up to \textbf{98.40\%} compared to direct camera offloading.
\end{abstract}



\maketitle

\section{Introduction and Related Work}
Advances in deep learning, hardware design, and modeling over the past decade have enabled the dream of autonomous vehicles (AVs) to become a reality. AVs are expected to improve road safety, passenger comfort, and mobility significantly. The core task of an AV is to perceive the state of the road and safely control the vehicle in place of a human driver. The difficulty in achieving this goal lies in the fact that road scenarios can be highly complex and dynamic, presenting a wide range of potential challenges and obstacles (e.g., rain, snow, construction zones, animals, etc.). To address this challenge, AV algorithms rely heavily on (i) large deep learning (DL) models to capture this high degree of complexity and (ii) high-performance edge hardware to reduce processing latency and ensure passenger safety at higher speeds.

As a result, AVs require significant computational power to operate reliably and safely in the real world. 
\blue{
However, as AV computing capabilities have scaled up, so have their power and energy requirements.
For example, the Nvidia Drive PX2, used in 2016-2018 Tesla models for their Autopilot system \cite{px2tesla}, can achieve 12 Tera Operations Per Second (TOPS) with a Thermal Design Power (TDP) of 250 Watts (W). Following the PX2 was the Nvidia AGX Pegasus, which was built for level 5 autonomy; it can achieve 320 TOPS with a TDP of 500W \cite{Oh2017pegasus}. What's more, the next generation hardware platform using the Nvidia AGX Orin SoC is expected to be capable of 2000 TOPS with a TDP of 800W \cite{Abuelsamid2020orin}.
Although AV hardware platforms are becoming more efficient in terms of TOPS/W, the baseline energy demands continue to increase as more advanced DL models and hardware platforms are developed.
The increased power demands of these systems also increase the heating, ventilation, and air conditioning (HVAC) system's thermal load. The combined computational and thermal loads of these platforms can reduce an AV's driving range by up to 11.5\% \cite{lin2018architectural}, which is especially detrimental for electric vehicles due to their limited range and long recharge times.}

Researchers attempting to address this problem for AVs as well as other cyber-physical systems have proposed several approaches for reducing energy consumption, including application-specific hardware design, cloud/fog server offloading, or model simplification/pruning \cite{lin2018architectural, kong2020computation, pruning,vatanparvar2018design,al2015energy, odema2021lens}.
Although solutions like Application-Specific Integrated Circuits (ASICs) can reduce energy consumption through hardware optimization, they are prohibitively expensive to develop. Furthermore, with ASIC designs, all model specifications and contingencies need to be accounted for at design time, meaning there is little to no support for adding new features, fixing algorithmic errors, or modifying model architectures. Costly development stages will need to be repeated for every revision to the model.
The next logical choice is to attempt model simplification/pruning without changing hardware platforms; however, it is difficult to significantly reduce energy consumption by pruning without adversely affecting the AV's performance and safety. 
To address the limitations of the previous two approaches, some works propose offloading some or all AV tasks to the cloud for processing to reduce the energy consumption of the AV without changing the hardware or algorithms. Unfortunately, current offloading approaches have significant scalability and latency issues, as will be discussed in the next paragraph.
In contrast, we propose a cloud server offloading methodology that is efficient, safe, and practical for current networking infrastructure.

A na\"ive solution to the problem of edge energy consumption is to offload self-driving tasks to a cloud server or a Mobile Edge Computing (MEC) server \cite{feng2018mobile}. These `direct offloading' approaches involve sending images or sensor inputs directly to the server, which processes the data before returning the desired control outputs to the vehicle. 
However, the real-time latency constraints of autonomous driving and the limitations of current wireless network infrastructure significantly impact this solution's feasibility; \green{to drive and react effectively, AVs must be able to process each input within 100 milliseconds \cite{lin2018architectural}. This bound comes from the fact that the fastest attainable reaction by a human when driving falls within the range of 100-150 ms, meaning that for efficient AV navigation, AVs need to at least perform at the same level as the human driver counterpart.} Additionally, most real-world AVs, such as those from Tesla \cite{px2tesla}, Baidu Apollo \cite{baiduapollo}, and Argo AI \cite{argoai}, use multiple high-definition cameras and sensors and would require very high network bandwidths to offload data within the latency constraints. In some cases, the energy needed to transmit and receive data from the cloud server can even \textit{exceed} the energy consumed by edge-only processing. 
Together, these factors make direct offloading infeasible in most real-world autonomous driving scenarios. Currently, most of the literature has proposed solving this problem by improving network robustness and throughput via solutions such as 5G C-V2X \cite{papathanassiou2017cellular} and WAVE \cite{eichler2007performance}, or even by placing sensors on the roads themselves \cite{kong2020computation}. However, implementing these solutions would require significant investments in the networking infrastructure to become realistically feasible.

Several works have proposed methods for offloading some or all AV tasks. For example, \cite{cui2020offloading} proposed a technique for reducing AV processing latency by offloading sub-tasks of LiDAR SLAM to the cloud depending on network conditions. Although they demonstrate good performance, their approach is limited since it only considers LiDAR data, which is significantly smaller than camera data. Additionally, they developed a distributed SLAM algorithm that allowed task-level parallelism; this sort of optimization will need to be applied for every part of a modular AV pipeline and may not be applicable in some areas. 
In another work, \cite{sasaki2016vehicle} proposed an offloading strategy where computations are executed on either an MEC server or a cloud server depending on network conditions. However, their method requires all sensor input data and internal state information to be sent to the server for processing. Since they only evaluated a micro-car transmitting IMU data (position, velocity, yaw), their approach is not scalable to real-world AVs that would need to offload multiple high-definition camera inputs. 
The work in \cite{zhang2017optimal} proposes a hierarchical approach for offloading in which AVs can offload to road-side units (RSUs) when MEC servers are overloaded, but this work does not consider network bandwidth constraints.
Moreover, none of these works \cite{cui2020offloading, sasaki2016vehicle, zhang2017optimal} considered edge energy consumption in their evaluation, which significantly constrains direct offloading approaches.
The authors in \cite{zhang2016energy} evaluated the energy consumption for offloading to MEC servers; however, they do not assess this approach's practicality for large upload data sizes, which are typical for AVs with multiple high-resolution input cameras. 
In summary, the problem of offloading large data sizes while meeting latency and energy constraints on current network infrastructure is exceedingly challenging and is currently unsolved by existing methods. 

\subsection{Research Challenges}
For efficient AV offloading, the following key research challenges need to be addressed:

\begin{enumerate}
    \item Offloading AV tasks without exceeding safety-critical latency constraints or increasing AV energy consumption.
    \item Adapting AV deep learning architectures to support dynamic offloading depending on the corresponding network conditions.
    \item Developing a technique efficient enough to meet latency constraints with data from multiple high-definition camera inputs on current industry-standard AV hardware.
    \item Producing a cost-efficient, safe solution that can operate within the constraints of current networking infrastructure.
\end{enumerate}

Instead of altering the AV hardware or the communication network infrastructure, we propose SAGE: a methodology to significantly reduce the size of the data transmitted over the network and enable efficient computation offloading. By introducing a \textit{bottleneck} layer near the beginning of end-to-end DL control models, the size of the data uploaded to the cloud server is reduced significantly, allowing a large portion of the model computation to be offloaded to the server even at low network bandwidths. This benefit is especially valuable in multi-camera offloading due to the significant bandwidth requirements and edge energy consumption of multi-camera models. Furthermore, it was shown in \cite{marco1} that, with a particular training strategy, the model's performance after introducing the \textit{bottleneck} remains nearly the same.

\begin{figure}
\centering
\begin{subfigure}[b]{0.49\textwidth}
    \centering
    \includegraphics[width=\textwidth]{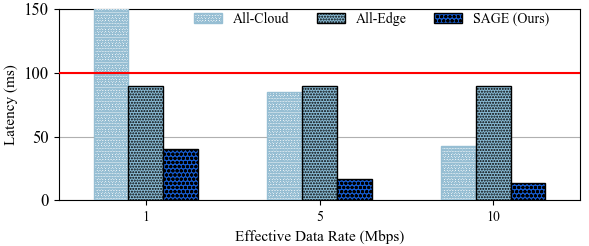}
\end{subfigure}
\begin{subfigure}[b]{0.49\textwidth}
    \centering
    \includegraphics[width=\textwidth]{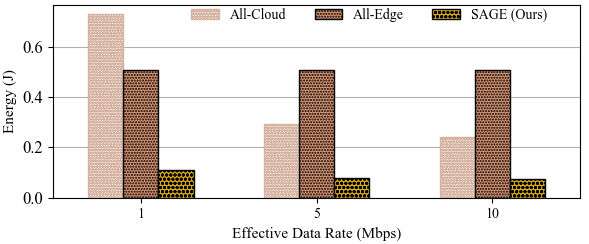}
\end{subfigure}
    \caption{A comparative analysis demonstrating the of the overall latency \textit{(left)} and energy consumption \textit{(right)} for the \textit{All-Cloud}, \textit{All-Edge}, and \textit{SAGE (Ours)} execution strategies given three typical effective data rate values attainable through a 4G LTE connection. The red line indicates the AV processing latency deadline of 100 ms.}
    \label{fig:motivation}
\end{figure}

\subsection{Motivational Example}
We provide a brief example to demonstrate the merit of our approach in Figure \ref{fig:motivation}. Here, we compare three possible execution strategies for an end-to-end AV control model: executing locally on the edge (\textit{All-Edge}), offloading the entirety of execution to the cloud (\textit{All-Cloud}), and our proposed split approach (SAGE). Our analysis is conducted at three distinct data rate values typical for 4G LTE connections, and the evaluations are performed on a Jetson TX2 for a ResNet-50 model \cite{resnet} adapted for end-to-end AV control. In terms of latency, it is clear that the \textit{All-Cloud} approach is impractical at low data rate values as it fails to meet the 100 ms processing latency constraint of the AV. On the other hand, performing all the processing locally in the \textit{All-Edge} approach keeps the latency unaffected by the state of the network. However, the downside is that the edge device is fully operational and consumes sizeable amounts of power for longer periods, leading to higher energy consumption in theory than the \textit{All-Cloud} approach at the more favorable data rates. SAGE offers to leverage the best of these both approaches. {\color{darkgreen} In brief, SAGE entails replacing an early computational block from the model architecture with a more efficient encoder-decoder-like structure. Then, this modified architecture is divided between the edge and cloud at the encoder output. The encoder, acting as a \textit{bottleneck}, projects the input data into a low-dimensional representation that is more suited to be transmitted to the cloud over the wireless medium. On the other hand, the decoder component is situated as part of the cloud to receive the encoder's output data and map it into a representation analogous to the output of the original computational block from the unaltered model architecture.} This structural modification results in significantly lower: (i) local execution latency than the \textit{All-Edge}, and (ii) transmission latency than the \textit{All-Cloud}. Moreover, these improvements are reflected in the energy consumption as the edge device is only required to perform computations for a much shorter interval within the 100 ms time window, which is beneficial for the edge device itself in terms of performance efficiency. \green{More details about the proposed SAGE methodology and how resource-efficiency is promoted across the edge and cloud while maintaining the same degree of accuracy shall be described in detail in Section \ref{sec:methodology}.}

\subsection{Novel Contributions}
Our paper presents the following contributions.

\begin{enumerate}
    \item We propose a novel split-network architecture methodology that allows for a significant reduction in the energy consumption of AV on-board processing units by dynamically offloading part of the model's computations to the cloud. 
    \item We demonstrate that introducing \textit{bottlenecks} into deep end-to-end AV control models reduces energy consumption significantly with little to no performance loss. 
    \item We show that SAGE reduces network throughput requirements significantly compared to conventional cloud server offloading techniques, enabling it to meet latency constraints even at low network bandwidths on 3G, 4G LTE, and WiFi\footnote{We did not evaluate 5G C-V2X and WAVE in this work because these technologies are currently not widespread and require significant infrastructure changes to be viable. Also, comparable real-world power models for 5G are not available yet in the literature.
    However, SAGE will be scalable to these emerging technologies.}.
    \item We demonstrate that SAGE is scalable to practical AV use cases by evaluating its performance with three high-definition camera inputs, typical for real-world AVs \cite{px2tesla,baiduapollo,argoai,hecker2018end}. 
    \item We demonstrate the practicality and feasibility of our technique by evaluating its performance on the Nvidia Jetson TX2, as well as the industry-standard Nvidia Drive PX 2 autonomous driving platform, used in all 2016-2018 Tesla models for their Autopilot system \cite{px2tesla}. 
\end{enumerate}

\subsection{Paper Organization}
The remainder of the paper is organized as follows. In Section \ref{sec:model}, we discuss our system model and problem formulation. In Section \ref{sec:methodology} we elaborate on SAGE's design methodology. In Sections \ref{sec:exp} and \ref{sec:discussion} we present and discuss our experimental results. Finally, in Section \ref{sec:conclusion} we present our conclusions.

\section{System Model}
\label{sec:model}
This section aims to provide a generalized model of how an AV edge device may complete processing a task either through local computation or collaboration with a cloud server. Mainly, the modeling comprises the communication and computation costs that the AV edge device would incur until the task is finished. Our model comprises a direct link between a vehicle $i$, requiring computation for its designated task, and a cloud server $j$ to whom tasks can be offloaded.

\subsection{Communication Model}
As the AV runtime optimization solution spans multiple levels in the system architectural hierarchy (i.e., edge and cloud), a communication model is needed to identify the cost of transferring data between entities of different levels. These costs can be represented through transmission latency and energy. More formally, the task to be offloaded can be represented as $t_{i} = \{a_{i}, b_{i}, c_{i}\}$, where $a_{i}$, $b_{i}$, and $c_{i}$ correspond to the size of data to be transmitted, size of data to be received back from the server, and the number of CPU cycles required to complete the task, respectively. \green{To estimate the communication overhead, we will need to determine the upload and download data rates, $r_{i,j}^{U}$ and $r_{i,j}^{D}$, experienced at vehicle $i$'s edge device when transmitting data to cloud server $j$. Although the data rate can be determined theoretically through Shannon's law, this resembles an optimistic estimate, not taking into consideration potential errors or packet losses. Instead, we are more interested in the '\textit{effective}' data rates by which we mean the actual data transfer speeds experienced at the edge device when accounting for errors and re-transmissions. These values can be measured at the target device and accordingly, the upload and download latencies can be given as:} 
\begin{align}
    T_{i, j}^{U} = \frac{a_{i}}{r_{i,j}^{U}}, \:\:\:\;\;\;
    T_{i, j}^{D} = \frac{b_{i}}{r_{i,j}^{D}}
\end{align}
Thus, the total communication overhead encountered by at vehicle $i$ in terms of latency and energy for offloading task execution to computing server $j$ is given by:
\begin{align}
T_{i, j}^{comm} &= T_{i, j}^{U} + T_{i, j}^{D} + T_{i, j}^{RTT} \\
E_{i, j}^{comm} &= p_{i}^{T}T_{i}^{U} + p_{i}^{R}T_{i}^{D} \label{eqn: ecomm}
\end{align}
where $p_{i}^{T}$, $p_{i}^{R}$ and $T_{i, j}^{RTT}$ represent vehicle $i$'s transmitting power, receiving power, and the round-trip time between vehicle $i$ and server $j$, respectively.

\subsection{Computation Model}

Assuming that any task requested by vehicle $i$ consists of several sequential sub-tasks, i.e., as in an end-to-end control pipeline or layers in a DL model, let $\mathbb{C}_{i} = \{c_{i1}, c_{i2}, ..., c_{iK}\}$ denote the set of K clock cycles required to execute each sub-task. Thus, potential execution times (local or remote) and the energy needed to execute sub-task $k$ locally are:
\begin{align}
    T_{ik}^{l} &= {\frac{c_{ik}}{f_{i}^{l}}} \\
    T_{ik}^{r} &= {\frac{c_{ik}}{f_{i}^{r}}} \\
    E_{ik}^{l} &= {\vartheta_{i}c_{ik}}
\end{align}

where $f_{i}^{l}$, $f_{i}^{r}$ and $\vartheta_{i}$ represent the operational frequency at vehicle $i$, operational frequency at the remote server, and a coefficient denoting energy consumed per CPU cycle at vehicle $i$. However, since offloading some or all sub-tasks is a viable option in this scheme, the total computational latency and energy consumption for vehicle $i$ can be written as:
\begin{align}
    T_{i}^{comp} &= \sum_{k=1}^{k_p}{T_{ik}^{l}} + \sum_{k=k_p+1}^{K}{T_{ik}^{r}} \label{eqn: tcomp}   \\
    E_{i}^{comp} &= \sum_{k=1}^{k_p}{E_{ik}^{l}} + E_{i}^{idle}(t) \label{eqn: ecomp}
\end{align}

where $k_p$ is the execution partitioning point after which execution is assigned to the remote server, and $E_{i}^{idle}(t)$ is the energy consumed by vehicle $i$ waiting for the remote server's results as a function of the idle time $t$. \green{Note that when $k_{p} = K$, $T_{i}^{comp}$ reflects the local execution case without any form of offloading as the second summation becomes an empty sum.}

\subsection{Problem Formulation}
\label{subsec:problem-formulation}
From the previous model derivations, the offloading problem for vehicle $i$ can be formulated as:
\green{
\begin{align}
\label{optimization-eqn}
\min\limits_{k_p}
\begin{split}
&w_{i}^{T}(\mathcal{I}(k_{p}\neq K)\times T_{i}^{comm} + T_{i}^{comp}) +  w_{i}^{E}(\mathcal{I}(k_{p}\neq K)\times E_{i}^{comm} + E_{i}^{comp}) 
\end{split}                                           \\
&\text{s.t.}
        \ (\mathcal{I}(k_{p}\neq K)\times T_{i}^{comm} + T_{i}^{comp}) <= 100\ ms      \notag 
        %
\end{align}
}
where $w_{i}^{T}$ and $w_{i}^{E}$ $\in [0, 1]$ represent user-defined weights associated with the latency and energy metrics, and \green{$\mathcal{I}(k_{p}\neq K)$ is an indicator function becoming 0 in the case of local execution.} As presented earlier, the 100 ms constraint is the window within which the AV must finish its processing task \cite{lin2018architectural}. Note that as $k_p$ varies, so will the values associated with the offloaded task $a_{i}$ and $c_{i}$.

\begin{figure}
\centering
    \includegraphics[width=\textwidth]{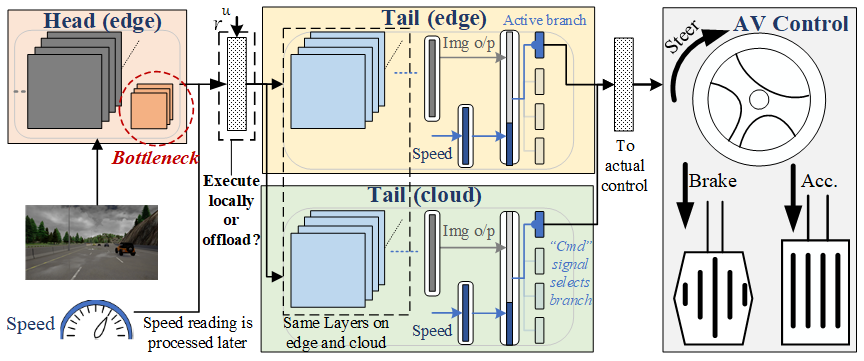}
    \caption{An illustration of how systems developed through SAGE support end-to-end AV control. The tail component is replicated on both the edge and the cloud. At runtime, a decision is to be made whether the tail should be executed locally or in the cloud. Final results are applied as inputs to the AV control system.}
    \label{fig:methodology}
\end{figure}

\section{SAGE Design Methodology}
\label{sec:methodology}
In this section, we present SAGE and discuss its building blocks in detail. Figure \ref{fig:methodology} illustrates how the final system developed through SAGE would support end-to-end AV control. The implemented DL model is divided into two components: (i) a \textit{head} deployed on the edge and (ii) a \textit{tail} which is replicated across both the edge and cloud. The \textit{head} component contains within its structure a \textit{bottleneck} layer, which represents an optimal offloading point compared to other options. Inputs to the model can come through a camera feed and sensory measurements (e.g., current speed). After the head portion executes at runtime, it is decided whether tail processing should be done locally or be delegated to the cloud depending on current network conditions. The \textit{tail} portion of each DL model contains the bulk of layers and outputs the control values to the AV.

\subsection{Perception}
Much like human beings, perception is concerned with how an AV interprets and understands events occurring in its surrounding environment. To enable perception, AVs are equipped with sensory capabilities to capture representative data from the environment. This data is then processed to extract a comprehensive understanding of the events unfolding around them. 
Contemporary AVs sense their environment via cameras, LiDAR, or radar equipment \cite{camera,radar,lidar}. After data acquisition, DL models process the data and estimate the course of action that the AV should take in the following time-step \cite{Codevilla2018}.
Without any loss of generality, our evaluations are based on the data-intensive image-based perception from a set of cameras capturing the AV's surroundings. To implement the perception pipeline, we utilize state-of-the-art DL model architectures, which are known to achieve high accuracy on image classification tasks, as baselines. This allows us to leverage these models' abilities to capture fine-grained features from images for processing camera data as part of an end-to-end AV control architecture. Mainly, we consider DenseNet-169 \cite{densenet}, ResNet-34 \cite{resnet} (used for end-to-end multi-camera AV control in \cite{hecker2018end}), ResNet-50 \cite{resnet}, and CarlaNet \cite{Codevilla2018} which is implemented specifically as an end-to-end AV control solution.

\begin{table}[ht]
    \centering
    \begin{tabular}{l c c c c}
    \hline
     & DenseNet-169 \cite{densenet} & ResNet-50 \cite{resnet} & ResNet-34 \cite{resnet} & CarlaNet \cite{Codevilla2018} \\
     \hline\hline
     Perception & 98.76\% & 97.74\% & 97.36\% & 59.64\% \\
     Imitation Learning (IL) & 1.24\% & 2.26\% & 2.64\% & 40.36\% \\
    \hline\hline
    Modified \textit{head} speedup & 80.11\% & 79.97\% & 67.25\% & 13.39\%\\
    Overall Model speedup & 27.01\% & 41.51\% & 34.39\% & 2.65\% \\
     \hline
    \end{tabular}

    \caption{Contribution of perception and IL components in terms of the total processing \textit{(top)}, and how modifying the \textit{head} components in SAGE speeds up model executions \textit{(bottom)}.}    
    \label{tab:layer breakdown}
\end{table}

\subsection{Imitation Learning for End-to-End Autonomous Vehicle Control}
Next, the baseline models must be adapted to predict AV control outputs from camera input data. This can be achieved by integrating an Imitation Learning (IL) component at the back-ends of the baseline models to enable them to mimic a human's behavior in regard to a particular task. In this context, the driving algorithm's core objective is to \textit{imitate} the vehicle control outputs (steering angle, brake pedal angle, and accelerator pedal angle) produced by a human driver for a given set of input images \cite{tampuu2020survey}. IL models are typically trained via supervised learning, where the goal is to map the input features captured at time-step $t$ to the corresponding human control output values. To effectuate the learning process, a loss function, e.g., Mean Absolute Error (MAE), is used to evaluate the difference between a model's predictions and the ground truth values.
\green{Take the baseline ResNet-50 for example, its vanilla network architecture constitutes five main convolutional blocks, representing the main perception component, followed by a final fully-connected layer for image classification tasks. To adapt the model for IL, we replace this fully-connected layer with an IL component developed for end-to-end AV control where the final layer has three separate neurons: one for each control output (steering, accelerator, brake). These outputs are used in both the loss function for MAE computation and controlling the vehicle during deployment.} We follow the IL implementation in \cite{Codevilla2018}  where firstly, the output from the preceding perception component and the corresponding pre-processed speed measurement at time-step $t$ are concatenated together as the input to the IL component.  Next, one of several processing branches is activated based on the driver's command value (e.g., navigation signal). This notion of branching is implemented to associate unique learning features with different driving intentions. For instance, the second branch can only be activated whenever the driver issues the \textit{"Turn right"} navigation signal because this branch's parameters were trained to take actions in anticipation of a right turn, dissimilar to what parameters in other branches learned. The outputs from the active branch are the ones that are directly applied to the AV control system at that particular time step $t$. 

To summarize, a baseline DL-based solution for AV control comprises (i) a perception module, (ii) a speed measurement processing unit, and (iii) an IL back-end. Henceforth, these DL models adapted for IL shall have \textit{"IL-"} preceding their original names, e.g., \textit{IL-ResNet-50}. Moreover, to give an idea of each component's contribution to the overall processing time, The upper part of Table \ref{tab:layer breakdown} shows how perception can be the most computationally-intensive component, especially when utilizing state-of-the-art image classification models. Note that the speed processing unit is executed concurrently with the perception module, which dominates their combined execution time. Thus, our structural modifications target the perception modules to maximize the performance impact.

\begin{figure}
\centering
    \includegraphics[width=\textwidth]{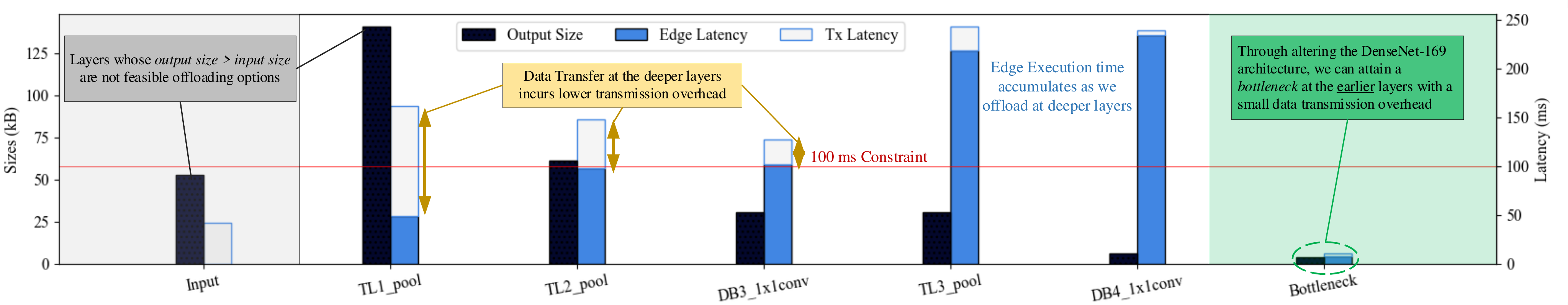}
    \caption{A comparison between offloading from: (i) the input, (ii) different perception layers in IL-DenseNet-169 \cite{densenet}, or (iii) the proposed \textit{bottleneck} layer. 
    The edge device is a Jetson TX2, the input is a $200\times88$ sized image, and the effective data rate is 10 Mbps. The IL-DenseNet-169 layers displayed are the ones which provided the smallest output data size. Note that the only options that do not violate the 100 ms latency constraint are offloading at the input or offloading at the \textit{bottleneck}, with the latter offering a $14\times$ overall speedup.}
    \label{fig:sizes}
\end{figure}

\subsection{Structural Alterations for Split Computing}
Deploying the AV control algorithm on the edge device is essential for such a mission-critical application. However, dynamically assigning some or all of the processing tasks to a more powerful cloud server if the wireless network conditions are favorable can lead to substantial latency and energy savings on the edge device. 
As was shown in the motivational example, directly offloading inputs to the cloud can be inefficient at sub-par network conditions: a significant communication overhead can arise from transmitting the raw input images, resulting in a poorer overall performance than that of local execution. Prior work in \cite{jpeg} tries to address this by compressing the input image before transmission, but accuracy degrades significantly. One alternative in \cite{neurosurgeon} proposes scanning each layer within the DL model to identify those which output smaller data sizes than the input as potential data offloading points in a split computing approach. However, this is dependent on each architecture's structure, deeming it ineffective for models that do not shrink data size enough.

A more tractable alternative is modifying the DL model structure by injecting a \textit{bottleneck} amongst the first few layers. This \textit{bottleneck} layer presents an optimal offloading point very early in the model because its structure is designed to output exceedingly small-sized data.
This idea is presented in \cite{marco1, marco2, marco3}, where it is implemented by initially dividing a DL model architecture into two sections: a \textit{head} and a \textit{tail}. The structure of the \textit{tail} remains unchanged. Whereas, a simpler more efficient version of the \textit{head} is constructed to mimic the functionality of the original \textit{head} section. The merits of constructing this new \textit{head} model are twofold. Firstly, the new \textit{head} is structurally more efficient to run than the original \textit{head} providing a local execution speedup, as illustrated in the lower part of Table \ref{tab:layer breakdown}. Secondly, the \textit{head} contains the \textit{bottleneck} operating as an encoder-decoder model rigorously transforming its input to lower dimensions (encoder) before raising the dimensionality at its output (decoder), making the encoder serviceable as an efficient data offloading layer. We follow the instructions provided in \cite{marco1} on how to design a new \textit{head} model with a \textit{bottleneck} from the original \textit{head}. Structurally, both the \textit{bottleneck}'s number of output channels and its preceding layers' complexity should be minimized. However, the modified model's accuracy still needs to be maintained by retraining the new head portion, as discussed in the following subsection. 
\green{The overhead for creating a bottleneck layer is analogous to that of creating a small deep learning model manually, which is represented through the human design effort of performing successive refinements in order to attain the desired degree of performance. The main difference is the requirement to have an encoder-decoder structure within the overall architecture to provide the efficient offloading point.}

As an example to demonstrate the efficacy of the \textit{bottleneck}, we compare performances at an injected \textit{bottleneck} in DenseNet-169 \cite{densenet} against offloading at the input or one of the six layers that provided the \textit{smallest} output (o/p) sizes in the original DenseNet-169. In this analysis, shown in Figure \ref{fig:sizes}, we assume a 10 Mbps connection using the $200\times88$ sized images from \cite{Codevilla2018} to calculate the overall latency with the Jetson TX2 as the edge device. Layers other than the \textit{bottleneck} are displayed according to their position within the DenseNet-169 architecture. The following trend can be observed from Figure \ref{fig:sizes}: as we go deeper in the network, the size of the output data at each layer decreases, reducing transmission overheads as we progress. However, to reach those lower layers, a considerable amount of execution needs to be performed locally. Thus, none of these deployment options outperform offloading at the input layer. On the contrary, injecting a \textit{bottleneck} early in the architecture decreases the output size to a value much smaller than the input, reducing transmission overhead. Moreover as a result of its early placement, intensive local processing is not required prior to the \textit{bottleneck} layer. All in all, offloading at the \textit{bottleneck} is $14\times$ faster than offloading at the input. This result would also be reflected in the energy consumption.
 From a formal perspective, the \textit{bottleneck} dominating all other offloading strategies transforms Equation \ref{optimization-eqn} into a runtime binary decision problem, in which either local execution is selected at extremely poor network conditions or offloading at the \textit{bottleneck} is chosen otherwise.

\begin{figure}
\centering
    \includegraphics[width=\textwidth]{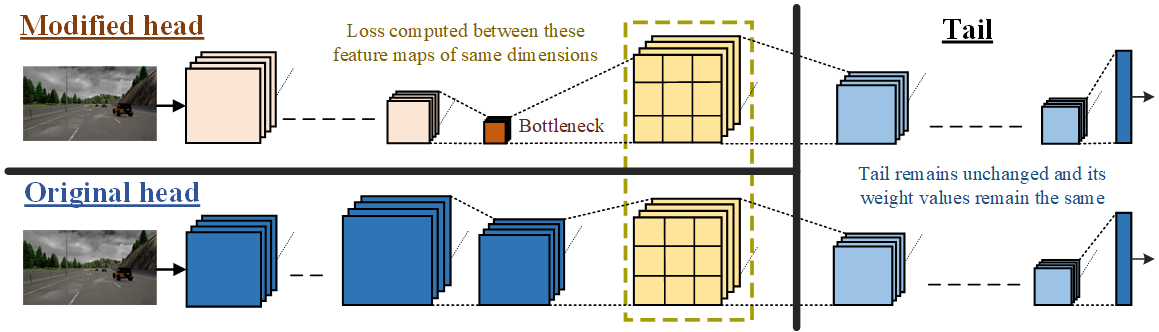}
    \caption{The head network distillation process to train the perception component with the \textit{bottleneck}. Note that in the final deployment, the \textit{bottleneck} layer is to be the last layer on the edge device.}
    \label{fig:HND}
\end{figure}

\subsection{Head Network Distillation (HND)}
\label{subsec:knowledge-distillation}
Knowledge Distillation (KD), presented in \cite{KD1, KD2}, has emerged as an effective training technique to render a compressed yet accurate version of a deeper, more complex neural network model. The main reason this technique came about is that shallower neural networks, when trained conventionally, achieve sub-par performance at many tasks compared to deeper networks. Hence, this technique aims to leverage the deeper network as a \textit{teacher} to \textit{distill} its acquired knowledge into the smaller \textit{student} model. Consequently, \textit{student} models trained through KD achieve superior performance relative to their traditionally-trained counterparts \cite{KD1}. This technique is advantageous when high-performance neural network solutions are needed for edge devices with limited resources. Formally, the \textit{student's} loss function, which is minimized during training, needs to incorporate a distillation component as follows: 
\begin{equation}
    \mathcal{L}_{student} = \alpha \mathcal{L}_{orig} + (1 - \alpha) \mathcal{L}_{KD}
\end{equation}
where $\mathcal{L}_{orig}$ is the conventional loss function using hard labels, whereas $\mathcal{L}_{KD}$ represents the KD loss component, which can be computed using KL divergence, L2 loss, or logits regression \cite{KD1}. Through providing a control variable $\alpha$, the effective weight of each loss component can be fine-tuned. This works because the \textit{student} is learning by minimizing the divergence from a vector of the \textit{teacher's} real values, rather than on a single label representation. Hence, the student becomes more capable of capturing the finer details of how the final decision was reached and attempts to learn a simple function to minimize the divergence from this vector of values, thus achieving better generalization. 

However, works in \cite{KD3, KD4} discuss how using more complex and accurate \textit{teacher} models makes training through KD for the \textit{student} models more challenging as a result of the capacity mismatch. In these scenarios, more training heuristics are introduced, and more restrictions are imposed on the structures of \textit{student} models. To avoid this in the context of the AV problem, KD is applied between the original and modified \textit{head} components rather than the entirety of models, making them the \textit{teacher} and \textit{student}, respectively. This process entails training the learnable parameters within the modified head model while maintaining the pre-trained tail parameters unchanged from the original model. Consequently, the loss component for the \textit{student-head} model can be computed using the sum of squared difference, as presented in \cite{marco1}:

\begin{equation}
    \mathcal{L}_{KD}(X) = \sum_{x \in X} ||s_{h}(x) - t_{h}(x)||^2
\end{equation}

where $s_{h}$ and $t_{h}$ represent the output vectors from the head portions of the \textit{student} and \textit{teacher} on an input $x$, respectively. For this loss function to be attainable, the final layers in both \textit{head} models must have the same dimensions. Figure \ref{fig:HND} illustrates the Head Network Distillation (HND) process. Note that although the loss function is computed between the final layers in the \textit{head} modules, in the final deployment, any layers succeeding the \textit{bottleneck} are deployed on the cloud. 

\begin{figure}
\centering
    \includegraphics[width=\textwidth]{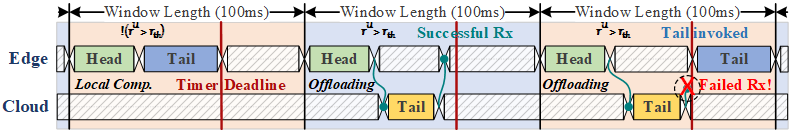}
    \caption{Three possible execution scenarios: local execution, successful offloading to the cloud, and unsuccessful offloading to the cloud which entails rolling back to edge computing}
    \label{fig:offloading-scenarios}
\end{figure}

\subsection{Offloading Strategy}
\label{subsec: offload}
After training the modified model using HND and deploying it across the edge and cloud, a runtime strategy must be implemented to determine, for each time step, whether to continue execution at the \textit{bottleneck} or delegate the remaining DL processing tasks to the cloud. The corresponding network conditions, mainly the effective upload data rate: $r_{i,j}^{U}$, govern this decision.
Note that the focus is on $r_{i,j}^{U}$ because the bulk of the data transmission (tens of kBs) occurs in the uplink, whereas merely the final values (in bytes) are sent through the downlink. So, the task here is to devise a policy based on a data rate threshold $r_{th}$ where:
\begin{enumerate}
    \item if ($r_{i,j}^{U}$ > $r_{th}$): the edge device offloads the result of computation at the \textit{bottleneck} to the cloud server where it is processed through the \textit{tail} part of the model before sending the resultant control inputs back to the edge device.
    \item \textit{else:} execution proceeds locally at the edge device. 
\end{enumerate}

To estimate $r_{th}$, the 100 ms constraint on AV processing, stated in Equation \ref{optimization-eqn}, must be considered, where all communication- and computation-related tasks must conclude within that time frame. Moreover, there have to be expected performance gains to justify the offloading decision. Therefore in our formulation, this decision is dependent on whether or not there exist potential energy reductions from offloading. Hence, we can denote $r_{th}$ as:
\begin{equation}
\begin{split}
    r_{th} = \frac{Upload\ Data\ Size}{100 - (T_{head}^{edge} + T_{tail}^{cloud} + T_{i,j}^{D} + T_{i,j}^{RTT})}\;\; s.t.\;\; (r_{th} > 0) \:and\: (E_{i}^{comm} + E_{idle}^{edge} < E_{tail}^{edge})
\label{eqn: decision}
\end{split}
\end{equation}

where $T_{head}^{edge}$ and $T_{i,j}^{RTT}$ represent the edge \textit{head} component's execution time and the round-trip time between the edge and cloud, respectively. The sum of $T_{tail}^{cloud}$ and $T^{D}$ represents the time the edge device is idle waiting for the cloud server to compute and transmit back the control inputs for the AV. The $r_{th} > 0$ restriction guarantees that the sum of the latency estimates in the denominator does not exceed the 100 ms time constraint. Furthermore, the sum of the energy required to offload the data at the \textit{bottleneck} and the idle energy consumption ($E_{i}^{comm} + E_{idle}^{edge}$) must be less than the energy required to execute the tail component of the model ($E_{tail}^{edge}$) in order to attain a \textit{beneficial} offload. Algorithm \ref{alg:optimization} demonstrates a runtime algorithm implementing this strategy. \green{We have built this algorithm to promote performance efficiency through offloading whenever the network conditions are benign. }\green{Note that \textit{\textbf{lines 12-14}} represent a fail-safe mechanism accounting for the network variability \textit{within} a single time window, where it is vital to keep room within the 100 ms time window to invoke local tail execution if the result has not received from the cloud within the expected time limit. This is guaranteed by starting a counter each time window that wakes the edge device to resume execution if the remaining time within the window is equivalent to that of the edge tail model.} Also, $T_{i,j}^{RTT}$ in our case is obtained through averaging multiple pings to a remote server, which accounted for $<10 ms$ overhead, however, this value may vary depending on the operational scenarios and the capabilities of network components involved in the communication link. Figure \ref{fig:offloading-scenarios} illustrates the three possible outcomes from our runtime strategy. \green{It should be noted that, since Algorithm \ref{alg:optimization} has a computational complexity of $\mathcal{O}$(1), its execution time is negligible compared to executing the DL models. As such, we excluded its execution time from our calculations. }

\begin{algorithm}
\caption{Runtime Energy Optimization Algorithm}
\DontPrintSemicolon
\label{alg:optimization}
	\KwIn{Upload Data Size, $T_{head}^{edge}, T_{tail}^{cloud}, E_{idle}^{edge}, E_{tail}^{edge}$}
		\For{\unboldmath{$each\ time\ step$} \boldmath{$t$}}{ 
		    Measure $r^{U}_{i,j}, r^{D}_{i,j},$ and $T_{i,j}^{RTT}$ \tcp*{obtain current network parameters}
		    Calculate $T_{i,j}^{D}, r_{th},$ and $E_{i}^{comm}$ \tcp*{using current network parameters}
		                ${x = edge\_head()}$    \tcp*{execute locally until bottleneck}	
		    \If{($r^{U} > r_{th}$) \textbf{and} ($r_{th} > 0$) \textbf{and} ($E_{i}^{comm} + E_{idle}^{edge}$ < $E_{tail}^{edge}$)}
		   {

		   \textbf{{$Tx\_data(x)$}} \tcp*{transmit bottleneck output to the cloud server} 
		   $Timer \gets reset()$ \tcp*{initialize timer}
		   $edge\_state \gets idle$ \tcp*{edge goes to idle mode} 
		   \If{rx\_event}{ 
		   $edge\_state \gets wakeup$ \tcp*{edge wakes up to receive server results} 
		   {$x = Rx\_data()$} \\
		   }
		   \ElseIf{$Timer > (100 - (T_{tail}^{edge} + \epsilon))$}
            {$edge\_state \gets wakeup$ \tcp*{edge wakes up to execute tail model} 
            {$x = edge\_tail(x)$} \\
		        }
		    }
		   \Else{
		   {$x = edge\_tail(x)$} \tcp*{execute tail model locally}

		   }
		   {$Input\_Control(x)$}  \tcp*{apply control values to the AV}

		}
	
\end{algorithm}

\section{Experiments}
\label{sec:exp}
\subsection{Experimental Setup}
We evaluate SAGE on two edge devices with significantly different computational capabilities: the Nvidia Jetson TX2 (TX2) and the industry-standard Nvidia DRIVE PX2 AutoChauffeur (PX2). The TX2 is capable of 1.33 TeraFlops (TFLOPS) while operating within a power budget of 15 Watts (W). The more powerful PX2 is designed for real-world autonomous driving use-cases and has been used in vehicles such as the Tesla Model S \cite{px2tesla}. It is capable of 8 TFLOPS within a power budget of 80 W. To serve as our cloud server, we used a Windows Desktop with an Nvidia 2080 Super, capable of 11.1 TFLOPS. It should be noted that, in a real-world deployment of SAGE, a more powerful cloud server could be used to increase the benefits of offloading.

In terms of the dataset, we use the CARLA conditional IL dataset from  \cite{Codevilla2018}. It contains RGB images in 200x88 resolution and control/sensor values extracted from the CARLA urban driving simulator \cite{dosovitskiy2017carla}. We used the image data as well as the steering, accelerator, brake, and navigational command information from the dataset for training and evaluating \green{the accuracy of both our original IL models as well as their bottlenecked counterparts}. We implement and train our models in PyTorch to assess the difference in error between the original and \textit{bottlenecked} models. 
\green{To evaluate the model latency and energy consumption ($T_{i}^{comp}$ and $E_{i}^{comp}$ from Equations \ref{eqn: tcomp} and \ref{eqn: ecomp}), we directly obtained the measurements through the Caffe model timing API for the TX2.}
For the PX2 and cloud server, we used Nvidia's TensorRT library to compile and optimize the models for the hardware. 
\green{TensorRT is designed to optimize the model architecture automatically (i.e., optimizing weights, parallelizing computations, combining redundant layers, etc.) to maximize inference performance on a given platform.}

In our experiments, we evaluated both 1-camera and 3-camera implementations of our IL models. Our 1-camera experiments aim to demonstrate SAGE for a low-cost AV implementation consisting of either a TX2 or PX2 as the edge device equipped with a single forward-facing camera. This implementation is practical for simple AV tasks such as adaptive cruise control, lane following, etc. Aligning with this goal, we evaluate the energy consumption and feasibility of SAGE with both low-resolution (88x200) and high-resolution (1280x720) camera data.

We also perform 3-camera experiments to demonstrate the feasibility of SAGE on more comprehensive AV hardware platforms. Multi-camera platforms are essential for real-world AV use cases such as urban/highway driving and point-to-point travel. Thus, we evaluated our IL models using three high-definition 1280x720 (720p) camera inputs on the PX2. Here, each model was modified to include three separate perception modules to process data from each camera. The outputs of the perception modules were then concatenated before being processed by the IL module, as was done in \cite{hecker2018end}. 

To evaluate the communication power cost needed in Equation \ref{eqn: ecomm}, we use the transmitting and receiving power models derived in \cite{close} for 3G, WiFi, and 4G LTE wireless technologies. Note that 5G energy evaluations are not available since we could not find any 5G-specific real-world power models in the literature as we found for the other technologies. In terms of the computation energy in Equation \ref{eqn: ecomp}, we leverage the onboard sensing circuits within the TX2 board for estimating the execution and idle powers, whereas we use an external power meter for the PX2. We assume no packet losses in our evaluations, and as mentioned, we demonstrate SAGE's feasibility with widespread and currently available network technologies.

\subsection{Performance Comparison of Original vs. Bottlenecked IL Models}
Recall that the \textit{bottleneck} acts as an encoder whose main purpose is to reduce the output data size to attain an efficient data transmission if needed. This data reduction is mainly achieved through reducing the number of \textit{output channels} at the \textit{bottleneck} layer (3 in the experiments). To give perspective, the \textit{output channels} for any layer in any of the original DL models discussed here before introducing our alterations is 32, meaning that there is an $\approx10\times$ reduction in output data size at least. To ensure that the introduction of a \textit{bottleneck} into our models does not impact their predictive performance, we evaluated the mean absolute error (MAE) of our models both with and without the bottleneck, shown in Table \ref{tab:performance_comparison}. In the cases with the bottleneck, we used HND to train the \textit{head} of the bottlenecked model to mimic the original model's \textit{head}, as described in Section \ref{subsec:knowledge-distillation}. The results clearly show that the bottlenecked models perform very similar to the original models, with only a slight increase in MAE. For context, an MAE increase of 0.01 corresponds to a 1\% increase in error between the model outputs and the outputs provided by the human driver. 

\begin{table}[ht]
    \centering
    \begin{tabular}{l l l l}
    \hline
    Model & \multicolumn{3}{c}{Mean Absolute Error (MAE)}\\\cline{2-4}
    & Steering & Accelerator & Brake\\\hline
    IL-DenseNet-169 & 0.0177 &0.0356 & 0.0129\\
    IL-DenseNet-169 w/HND & 0.0159 (-0.0018) & 0.0509 (+0.0153) & 0.0195 (+0.0066)\\\hline
    IL-ResNet-34 & 0.0259 & 0.0506 & 0.0199\\
    IL-ResNet-34 w/HND  & 0.0263 (+0.0004) & 0.0545 (+0.0039) & 0.0259 (+0.0060)\\\hline
    IL-ResNet-50 & 0.0260 & 0.0514 & 0.0180\\
    IL-ResNet-50 w/HND & 0.0266 (+0.0006) & 0.0601 (+0.0087) & 0.0330 (+0.0150)\\\hline 
    IL-CarlaNet & 0.0259 & 0.0546 & 0.0228\\
    IL-CarlaNet w/HND & 0.0204 (-0.0055) & 0.0589 (+0.0043) & 0.0326 (+0.0098) \\\hline 
    \end{tabular}
    \vspace{5pt}
    \caption{Comparison between the original IL models and their modified counterparts with \textit{bottlenecks} after HND. Values in parentheses are the differences in error between the models.}
    \label{tab:performance_comparison}
\end{table}

\subsection{Power, Energy, and Latency Evaluation on Hardware}
From this point onwards, all IL models referred to are with \textit{bottleneck} layers added.
In Table \ref{tab:hardware_energy_latency}, we compare the power consumption, energy consumption, and latency of different parts of the IL models on each hardware platform. 
By comparing the end-to-end (E2E) latency of the edge devices with the edge head latency and server tail latency, we see that offloading at the head provides ample time to account for network transmission latency. Furthermore, the table shows a significant energy reduction when processing the head model instead of the entire model end-to-end. These metrics illustrate the feasibility and potential benefits of our model.

\begin{table}[ht]
    \centering
    \begin{tabular}{@{\extracolsep{2pt}}p{62pt} p{27pt} p{20pt} p{20pt} p{20pt} p{28pt} p{28pt} p{28pt} p{26pt} p{28pt}@{}}
        \hline
        \multirow{2}{*}{Network} & \multirow{2}{*}{Device} & \multicolumn{3}{c}{Power (W)} & \multicolumn{3}{c}{Latency (ms)} & \multicolumn{2}{c}{Energy (J)} \\\cline{3-5}\cline{6-8}\cline{9-10}
        & & E2E & Head & Idle & E2E & Head & Tail & E2E & Head \\\hline
        \multirow{3}{*}{IL-DenseNet-169} 
        & Server & -- & -- & -- & -- & -- & \textbf{2.238} & -- & --\\
        & TX2 & 5.446&5.430&1.659&215.543&\textbf{8.043}&207.5&1.1740&\textbf{0.0437}\\
        & PX2 & 43.58&47.42&40.23&10.420&\textbf{1.112}&9.308&0.4541&\textbf{0.0527}\\ \hline
        \multirow{3}{*}{IL-ResNet-34} 
        & Server &--&--&--&--&--&\textbf{0.572}&--&--\\
        & TX2 & 5.95&5.221&1.659&65.560&\textbf{11.612}&53.948&0.3901&\textbf{0.0606}\\
        & PX2 & 46.99&47.51&40.23&4.534&\textbf{0.695}&3.839&0.2131&\textbf{0.0330}\\ \hline
        \multirow{3}{*}{IL-ResNet-50} 
        & Server & --&--&--&--&--&\textbf{0.607}&--&--\\
        & TX2 &5.682&5.415&1.659&89.231&\textbf{10.432}&78.799&0.5070&\textbf{0.0565}\\
        & PX2 & 46.89&47.17&40.23&7.413&\textbf{1.195}&6.218&0.3476&\textbf{0.056}4\\ \hline
        \multirow{3}{*}{IL-CarlaNet} 
        & Server & --&--&--&--&--&\textbf{0.188}&--&--\\
        & TX2 & 5.391&5.039&1.659&28.795&\textbf{8.727}&20.068&0.1552&\textbf{0.0440}\\
        & PX2 &45.54&46.33&40.23&1.659&\textbf{0.593}&1.066&0.0756&\textbf{0.0275}\\ \hline
    \end{tabular}
    \vspace{5pt}
    \caption{Hardware performance metrics for processing one 88x200 camera input. E2E = processing the entire model end-to-end on the edge device. Cloud server power/energy is ignored because this is not a constraint.} 
    \label{tab:hardware_energy_latency}
\end{table}

\subsection{Offloading Evaluation}
\label{subsec:1camera}

\subsubsection{Low Resolution}
\begin{figure}
\centering
    \includegraphics[clip, trim=140 178 140 178, width=\textwidth]{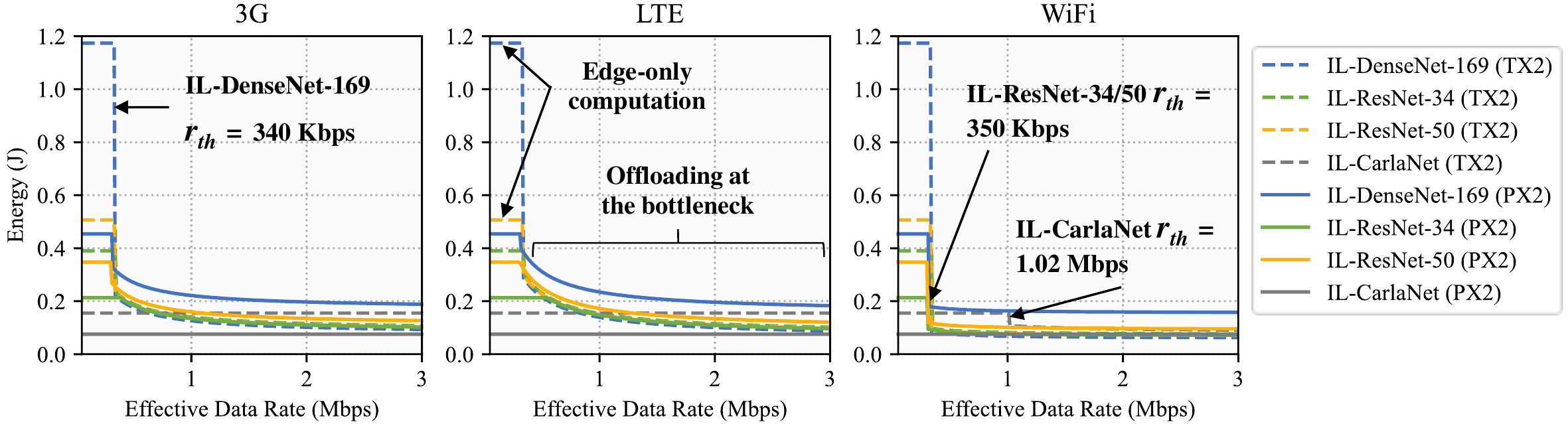}
     \caption{Energy consumption of IL models developed through SAGE while processing a single 88x200 camera input at different data rates with 3G, 4G LTE, and WiFi. The transition point in each line occurs at $r_{th}$, which is when offloading begins at the \textit{bottleneck}. Before this point, the energy consumption for the edge-only processing is ($E_{head}^{edge} + E_{tail}^{edge}$). After this point, the energy consumption is calculated as ($E_{head}^{edge} + E_i^{comm} + E_{idle}^{edge}$).}
     \label{fig:energy-1camera}
\end{figure}

\begin{figure}
\centering
    \includegraphics[clip, trim=250 197 250 196, width=.75\textwidth]{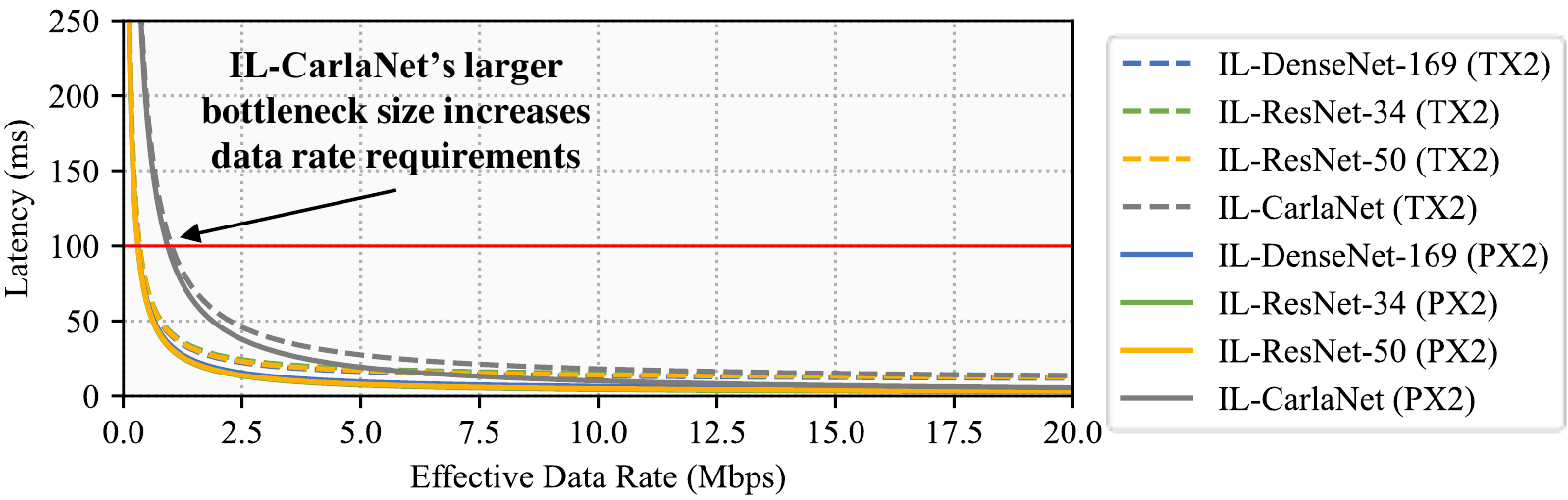}
    \caption{End-to-end latency of each model for  offloading at the \textit{bottleneck} for an AV with a single 88x200 camera input. The end-to-end latency includes edge head processing latency, wireless network latency, and server processing latency at various network data rates. The red line indicates the 100 ms latency constraint}
    \label{fig:latency-1camera}
\end{figure}

In Figures \ref{fig:energy-1camera} and \ref{fig:latency-1camera}, we show results from evaluating models implemented through SAGE with a single 200x88 resolution camera input using the TX2 and the PX2. 

Figure \ref{fig:energy-1camera} shows the energy consumption of each IL model with each technology type at different values of effective data rate $r^{U}$. Recall that these values are obtained based on the offloading strategy in Section \ref{subsec: offload}, where it is only feasible to offload when ($r_{i,j}^{U}$ > $r_{th}$) and ($E_{i}^{comm} + E_{idle}^{edge}$ < $E_{tail}^{edge}$). For each model, observe the sharp change in Figure \ref{fig:energy-1camera} at $r_{th}$ where the model switches from edge-only computation to cloud offloading. For IL-DenseNet-169, IL-ResNet-34, and IL-ResNet-50, this switching point occurs at approximately 350-400 Kbps on both the TX2 and PX2.

Although offloading can meet the latency constraint for some $r^{U}$ values, the energy consumed by the networking components must still be considered. As shown in Figure \ref{fig:latency-1camera}, IL-CarlaNet can feasibly offload at 1 Mbps on both devices, but on 3G and 4G LTE, offloading consumes more power than edge-only computation.  
Thus, we only consider $r^{u}$ values which are greater than $r_{th}$, at which offloading \textit{saves} energy on the edge device compared to edge-only processing.
For IL-CarlaNet on the TX2, $r_{th}$ is 7.7 Mbps on 3G, 3.62 Mbps on 4G LTE, and 1.02 Mbps on WiFi. This is likely because IL-CarlaNet has a larger data size at the \textit{bottleneck} than the other models, increasing communication latency and energy.
Interestingly, on the PX2, IL-CarlaNet's $r_{th}$ is 13.66 Mbps for WiFi and there is no 3G or 4G LTE $r_{th}$ under 100 Mbps for IL-CarlaNet that saves energy compared to edge-only computation. This is likely a result of the data size and the fact that IL-CarlaNet is a very small model and the PX2 has a moderately high idle power consumption (40.23W), meaning that offloading would consume more power than simply running on the edge. Since  IL-DenseNet-169, IL-ResNet-34, and IL-ResNet-50, are larger models, there is a clear benefit to offloading. Thus, the $r_{th}$ remains at 320-390 Kbps. The only exception is IL-ResNet-34 using LTE on the PX2, which has an $r_{th}$ of 550 Kbps, likely due to the efficiency of the PX2 compared to its idle power consumption and the network latency at this data rate.

Overall, when offloading at the $r_{th}$ for each model and technology, the TX2 and PX2 consume an average of \textbf{49.78\%} and \textbf{22.48\%} less energy, respectively, compared to edge-only computation. Interestingly, when running edge-only, the PX2 consumes half as much energy as the TX2; however, when both offload at $r_{th}$, the PX2 consumes $\approx25\%$ more energy than the TX2. This is likely because the network latency outweighs the efficiency benefit of the PX2 at these low throughputs. Regardless, both devices significantly reduce edge energy consumption by offloading.

For all models except IL-CarlaNet, the $r_{th}$ is well within the operating range for all three network technology types. Figure \ref{fig:latency-1camera} clearly shows that all models can meet the deadline of 100 ms with network throughputs as low as 320 Kbps. Above 15 Mbps, the benefit of higher data rates is minimal for this data size.

\subsubsection{High Resolution}
\begin{figure}
\centering
    \includegraphics[clip,trim=140 178 140 178, width=\textwidth]{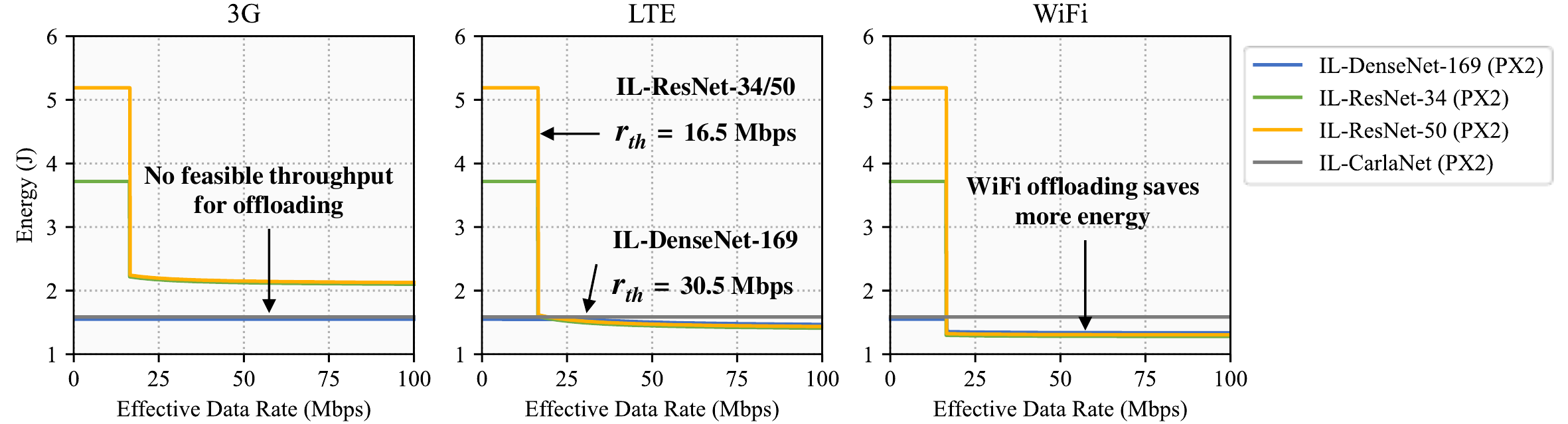}
    \caption{Energy consumption of each model for processing a single 1280x720 (720p) camera input}
    \label{fig:energy-1camera-720}
\end{figure}

\begin{figure}
\centering
    \includegraphics[clip, trim=250 197 250 196, width=.75\textwidth]{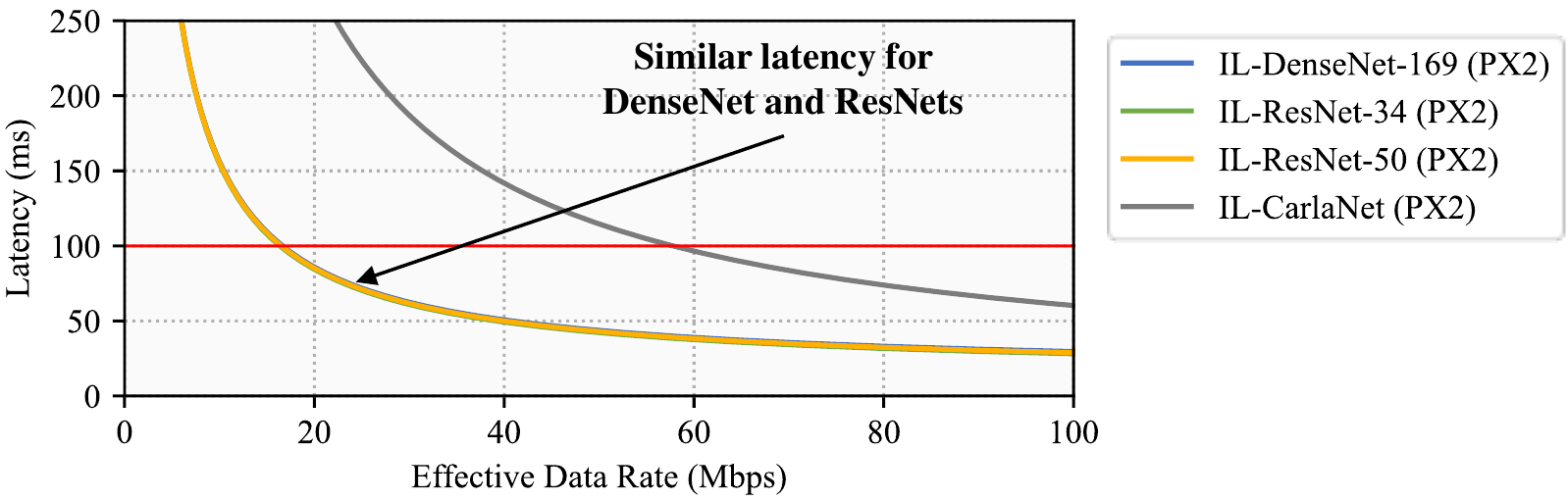}
    \caption{End-to-end latency of each model for offloading at the \textit{bottleneck} at different network data rates for an AV with a single 1280x720 (720p) camera input.}
    \label{fig:latency-1camera-720}
\end{figure}

The previous experiment demonstrated that our approach is feasible and has significant benefits for low-resolution camera data. However, real-world AVs use high-definition cameras to improve perception performance and safety \cite{baiduapollo, px2tesla, argoai}. To emulate this application, we evaluate SAGE on camera data with a 1280x720 (720p) resolution, the resolution used for Tesla Autopilot 2.0 systems. We only assessed the PX2 on this application since it is infeasible for the TX2 to meet the deadline of 100 ms with this input size even when running on the edge only.
The results of this experiment are shown in Figures \ref{fig:energy-1camera-720} and \ref{fig:latency-1camera-720}. 

The larger input image size increases model sizes and data sizes at the \textit{bottleneck} ($59\times$ larger for IL-CarlaNet and $47\times$ larger for all other models), increasing the edge processing latency, communication latency, and energy consumption significantly. This change is reflected in the figures, as IL-ResNet-34 and IL-ResNet-50 have an $r_{th}$ of $\approx16.5$ Mbps. This data rate is well within the normal operating ranges of 4G LTE and WiFi connections. 
IL-DenseNet-169 has $r_{th}$ values of 30.53 Mbps and 16.65 Mbps on 4G LTE and WiFi, respectively, but does not have a practical $r_{th}$ under 100 Mbps for 3G. This is likely because 3G consumes significantly more energy to upload data than 4G LTE and WiFi. Also, TensorRT better optimized the IL-DenseNet-169 model since it consists of a large number of relatively small layers, reducing its energy consumption significantly compared to IL-ResNet-34 and IL-ResNet-50. This reduction decreases the potential benefits of offloading in this case. 
Compared to edge-only processing, offloading the models at $r_{th}$ with 3G, 4G LTE, and WiFi reduces edge energy consumption by \textbf{48.54\%}, \textbf{41.96\%}, and \textbf{50.72\%}, respectively. With high-resolution data, the energy consumption benefit is more than double that of offloading low-resolution data, indicating that offloading is more beneficial for large, demanding edge models.

Since the data size at the IL-CarlaNet model's \textit{bottleneck} is $3.86\times$ larger than that of other models, it requires a higher throughput (57.8 Mbps) than the other models to meet the deadline. Also, IL-CarlaNet's small model size reduces its edge energy consumption, meaning that the communication energy consumption and idle power consumption could outweigh any potential savings. We found no $r_{th}$ below 100 Mbps for any networking technology that reduces the energy consumption of IL-CarlaNet below that of edge-only processing.

\subsection{Multi-Camera Evaluation}
\label{subsec:3camera}
State-of-the-art AVs use multiple high-definition cameras to capture more information about the vehicle's surroundings to improve decision-making, control, and safety \cite{hecker2018end, px2tesla, baiduapollo, argoai}. This problem is highly demanding in terms of energy consumption and network connectivity since the latency constraint remains the same at 100 ms despite the significant increase in input and model size. To evaluate SAGE on this application, we provide three 720p camera inputs to our models. 

We adapt our models for this task by replicating the original 720p perception pipelines to form three parallel perception pipelines (one for each camera input). The outputs of these pipelines are then concatenated and passed to the IL portion of each model. Consequently, each of the parallel perception pipelines contains one \textit{bottleneck} layer from which data can be offloaded. During offloading, we assume the data at all three \textit{bottlenecks} are sent to the cloud simultaneously. To reduce the maximum throughput requirement in this application, we quantize the values at the \textit{bottleneck} from 32-bit precision to either 16-bit or 8-bit precision before transmission.  
We tested IL-DenseNet-169 with quantizations of 16-bits and 8-bits at the \textit{bottleneck} layer and found that the average difference in MAE compared to the original is just \num{1.6e-10}, which is imperceptible.
Thus, with 16-bit and 8-bit quantization, we reduce our throughput requirements by 50\% and 75\%, respectively, while having a negligible effect on performance. 
Once again, we only evaluate the PX2 in this application since the TX2 cannot meet the deadline of 100 ms with the 3-camera models.

\begin{figure}
\centering
    \includegraphics[clip,trim=125 178 125 178, width=\textwidth]{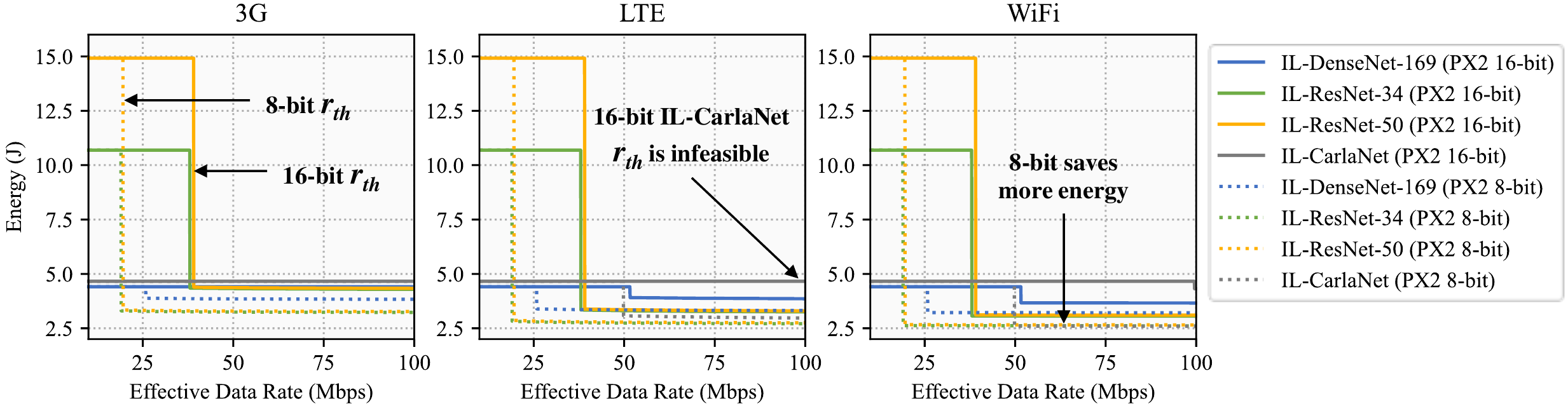}
    \caption{Energy consumption of each model for processing three 1280x720 (720p) camera inputs. Results are shown for both 16-bit quantization and 8-bit quantization at the \textit{bottleneck}.}
    \label{fig:energy-3camera-720}
\end{figure}

\begin{figure}
\centering
    \includegraphics[clip, trim=230 197 230 196, width=.75\textwidth]{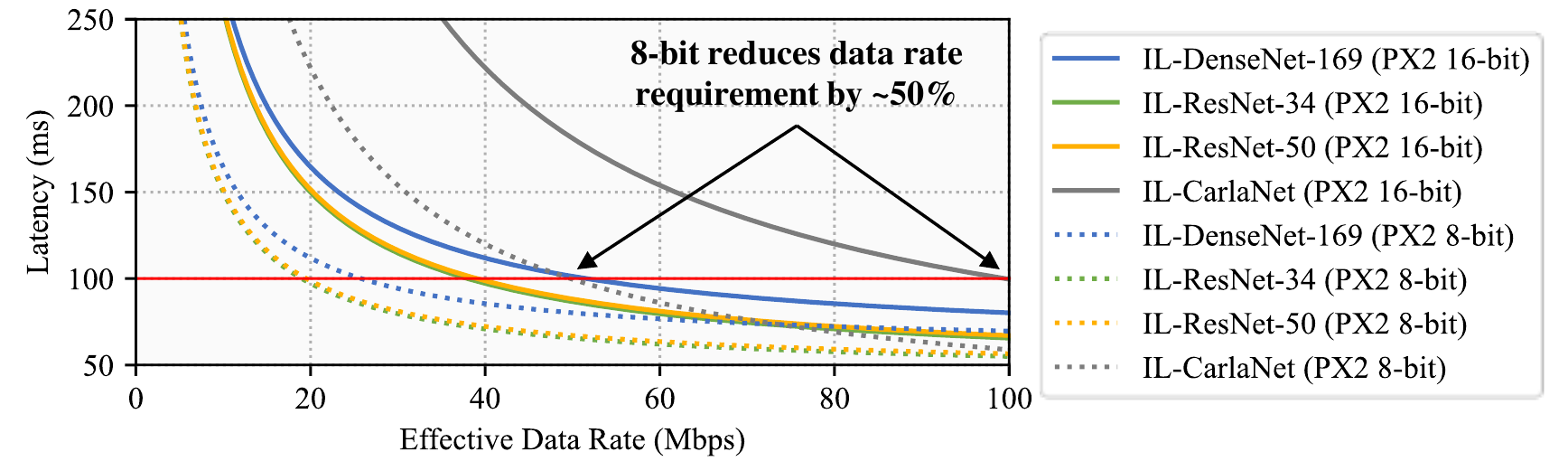}
    \caption{End-to-end latency of each model for offloading at the \textit{bottleneck} at different network data rates for an AV with three 1280x720 (720p) camera inputs. Results are shown for both 16-bit quantization and 8-bit quantization at the \textit{bottleneck}.}
    \label{fig:latency-3camera-720}
\end{figure}

In this application, all-cloud offloading approaches are entirely infeasible. Given that the input data size (three 720p images) is 8.29 MB total, they would require a minimum throughput of 664 Mbps to meet the 100 ms deadline. In contrast, the data size offloaded by our model with 16-bit \textit{bottleneck} quantization is only 264 KB ($31\times$ smaller); with 8-bit quantization, this drops to 132 KB ($62\times$ smaller).
In our experiments, we find that our approach is feasible at throughputs easily achievable by WiFi and 4G LTE. Our experimental results are shown in Figures \ref{fig:energy-3camera-720} and \ref{fig:latency-3camera-720}.

As shown in Figures \ref{fig:latency-3camera-720} and \ref{fig:energy-3camera-720}, with 16-bit quantization, IL-DenseNet-169, IL-ResNet-34, and IL-ResNet-50 can all offload at $r_{th}$ values of 51.57 Mbps, 37.98 Mbps, and 39.05 Mbps, respectively. With 8-bit quantization, these $r_{th}$ values drop to 25.79 Mbps, 18.99 Mbps, and 19.53 Mbps, respectively.
With 8-bit quantization, most 4G LTE and WiFi connections can easily support the $r_{th}$ data rates. 
Regarding 16-bit quantization, good quality 4G LTE and most WiFi connections should be able to support the $r_{th}$ data rates \cite{imoize2020analysis}.
On 4G LTE and WiFi, these models consume \textbf{52.67\%} and \textbf{50.40\%} less energy, respectively, by offloading at their $r_{th}$ throughputs. The energy reduction is much more significant for IL-ResNet-34 and IL-ResNet-50 than IL-DenseNet-169, which we again attribute to TensorRT's model optimizations.
It should be noted that, during offloading, all models appear to have very similar energy consumption. Practically, this means that an AV can run much larger models (e.g., use IL-ResNet-50 instead of IL-ResNet-34) without much difference in energy consumption provided a network connection with a data rate greater than $r_{th}$ is available most of the time. 

Once again, there is little benefit for offloading IL-CarlaNet due to the larger data size at the \textit{bottleneck} (1.01 MB) and the relatively low energy consumption of the model running on the edge. With 16-bit quantization, IL-CarlaNet only saves energy on WiFi at a data rate above 99.51 Mbps. However, with 8-bit quantization, offloading becomes feasible for both 4G LTE and WiFi at 49.76 Mbps. Since IL-CarlaNet is a relatively small model, it may be better to run it on the edge device most of the time and only offload on WiFi when network throughput is high.

\section{Discussion}
\label{sec:discussion}
In this section, we discuss our key findings from our experiments as well as the limitations, feasibility, and cost of SAGE. We also discuss future research directions.

\subsection{Overall Findings}
We found that SAGE was feasible for most IL models for all hardware configurations. By offloading at $r_{th}$, SAGE reduced edge device energy consumption by \textbf{36.05\%} with one low resolution camera, \textbf{47.07\%} with one high-resolution camera, and \textbf{55.66\%} with three high-resolution cameras. More energy could be saved by offloading at throughputs higher than $r_{th}$ when possible. Additionally, our results indicate that SAGE saves more energy by offloading when input data sizes are larger (i.e., when using more cameras or higher resolutions). SAGE also reduces upload data size by \textbf{96.81\%} and \textbf{98.40\%} with 16-bit and 8-bit quantization, respectively, compared to directly offloading three 720p camera inputs.
Besides, we found that our introduction of \textit{bottleneck} layers only increased mean error by $\approx1\%$ and quantization had a negligible effect on error, meaning that SAGE could be scaled to even higher camera resolutions easily.

\subsection{Limitations}
In our experiments, we found that our offloading methodology was not particularly effective for IL-CarlaNet. With low-resolution data, it required a significantly higher $r_{th}$ to provide a benefit than the other models; with high-resolution data and multiple cameras, there was no $r_{th}$ below 100 Mbps that reduced energy consumption. In its current form, SAGE may not present useful offloading for small models and models with a proportionally large \textit{bottleneck} size due to the increased energy cost of transmitting and receiving data compared to just running the entire model on the edge. 

\green{Additionally, although the 100 ms represents a reasonable worst-case bound, the current industry standard for real-time video processing is 30 frames/second, meaning that practically, the bound for completing the AV prediction task can be even tighter reaching $\approx$ 33 ms. From our experimental analysis, SAGE can meet this constraint when offloading the quantized version of the single full HD image data transformation. However, it fails to satisfy this requirement in the case of 3 HD camera inputs. Thus, experimentation with respect to AV industry-standard hardware and 5G wireless technology can provide a fair assessment of SAGE's capability to meet these tighter bounds.}

\blue{Although our methodology has shown promise in terms of improving the overall performance efficiency, several other factors can impact the extent of this improvement given some real-world situations. It is possible that channel contention between users, packet loss, and channel coherence issues related to vehicle speed and environmental conditions could limit the benefits of our methodology. These effects are difficult to simulate accurately, so real-world experiments are still needed to gauge the energy savings offered by our methodology in these situations.}

Lastly, we did not evaluate our approach on modular pipelines. However, since modular pipelines' perception modules generate the most latency \cite{lin2018architectural}, SAGE could be directly applied to these modules to achieve similar energy benefits. 
\blue{AV hardware platforms also handle other tasks such as route planning and user interfaces, but these applications constitute a minute part of the overall AV driving system.}
\blue{\cite{lin2018architectural} has shown that the object detection, tracking, and localization modules ( i.e., components of the modular version of the perception pipeline) comprise over 98\% of the total computation, consuming 1.99 J per input.} This proportion is very similar to the results we show in Table \ref{tab:layer breakdown}. Based on our energy savings with 3-camera offloading, if we introduce a \textit{bottleneck} to the object detection module and offload the remaining modules to the cloud, we could reduce energy consumption from 1.99 J to 0.896 J, a savings of 55\%. 

\subsection{Practicality and Cost}
Since SAGE does not require any hardware modifications to the AV or network infrastructure, it is much more cost-efficient and flexible than other solutions such as ASIC design or 5G C-V2X/WAVE installation. The only added costs are those associated with hosting a cloud server to run the offloaded models. However, we demonstrated SAGE's feasibility with a Desktop PC as the cloud server, so hosting similar hardware in the cloud would likely be inexpensive. These costs could even be passed on to consumers, where a vehicle owner could elect to extend their AV driving range by paying for an offloading service as proposed in \cite{zhang2017optimal}. Compared to direct offloading, SAGE has significantly lower throughput requirements, making it much more practical for real-world deployment with the current networking infrastructure.

\subsection{Future Work}
\green{In this work, we demonstrated the performance benefits attainable through the SAGE methodology over two NVIDIA hardware platforms, JETSON TX2 and DRIVE PX2. Although our approach is platform-agnostic, we intend to apply SAGE in our future works on different target hardware with different capabilities, like the high performance inference Neural Processing Units (NPUs) developed by ARM \cite{arm-npu}}.
\green{To ensure that our methodology does not introduce additional safety risks, it would also be prudent to evaluate each model on closed-loop evaluations in future work, such as judging each model's success rate at driving point-to-point in a simulator as in other works \cite{Codevilla2018, codevilla2018offline, codevilla2019exploring,dosovitskiy2017carla}.}
Moreover, even though we demonstrated the merit of SAGE using the current prevalent network technologies, this research area is still relatively new, and problems such as energy optimization with multiple servers, modular AV architectures, and 5G networks remain unstudied.
For example, SAGE can be adapted to address a multi-MEC server problem context. In this case, the action-space would expand from the AVs' perspective, for they would not only need to make an offloading decision each time step, but also identify which server should be selected for data transfer and task delegation. This would also entail additional dynamic factors to be considered, such as each server's load. Hence, a more sophisticated approach, like reinforcement learning \cite{RL1}, would need to be applied to solve the problem each time-step, in which previous connection experiences with the various servers could be leveraged through an in-place policy to guide the MEC server selection. 
These problems are left to be addressed in future works.

\section{Conclusion}
\label{sec:conclusion}
Designing AV control algorithms that are both safe and energy-efficient is a complex challenge that cannot be practically solved using simple direct offloading strategies. In this work, we proposed SAGE: a methodology for splitting the computation of IL end-to-end control models between the edge and the cloud while minimizing network throughput requirements by adding \textit{bottleneck} layers to the models. We evaluate SAGE on both large and small IL models and show that adding \textit{bottleneck} layers only results in a minor performance impact.
Our experiments demonstrate that SAGE reduces the edge energy consumption of IL end-to-end control algorithms with both low-resolution and high-resolution camera data by \textbf{36.13\%} and \textbf{47.07\%}, respectively. 
Additionally, we show that SAGE is scalable to AVs that use three high-definition camera inputs, reducing energy consumption by \textbf{55.66\%}, and can be practically implemented using current state-of-the-art AV hardware (PX2) and networking infrastructure (3G, 4G LTE, and WiFi).
On all three applications, we demonstrate that the IL models can be offloaded at effective data rates that are well within the constraints of current network infrastructure while still meeting AV latency deadlines. 
We also find that the throughput requirements for offloading reduce by 50\% and 75\% when quantizing the \textit{bottleneck} output to 16-bits and 8-bits, respectively, with a negligible change in model performance. 
Overall, we show that SAGE is practical for real-world, end-to-end control applications and can significantly curtail AV energy consumption.

\begin{acks}
This work was partially supported by the National Science Foundation (NSF) under award CMMI-1739503 and Graduate Assistance in Areas of National Need (GAANN) under award P200A180052. 
Any opinions, findings, conclusions, or recommendations expressed in this paper are those of the authors and do not necessarily reflect the views of the funding agency.
\end{acks}

\bibliographystyle{ACM-Reference-Format}
\bibliography{bibliography}


\begin{thebibliography}{42}


\ifx \showCODEN    \undefined \def \showCODEN     #1{\unskip}     \fi
\ifx \showDOI      \undefined \def \showDOI       #1{#1}\fi
\ifx \showISBNx    \undefined \def \showISBNx     #1{\unskip}     \fi
\ifx \showISBNxiii \undefined \def \showISBNxiii  #1{\unskip}     \fi
\ifx \showISSN     \undefined \def \showISSN      #1{\unskip}     \fi
\ifx \showLCCN     \undefined \def \showLCCN      #1{\unskip}     \fi
\ifx \shownote     \undefined \def \shownote      #1{#1}          \fi
\ifx \showarticletitle \undefined \def \showarticletitle #1{#1}   \fi
\ifx \showURL      \undefined \def \showURL       {\relax}        \fi
\providecommand\bibfield[2]{#2}
\providecommand\bibinfo[2]{#2}
\providecommand\natexlab[1]{#1}
\providecommand\showeprint[2][]{arXiv:#2}

\bibitem[\protect\citeauthoryear{??}{px2}{2016}]%
        {px2tesla}
 \bibinfo{year}{2016}\natexlab{}.
\newblock \bibinfo{title}{{All new Teslas are equipped with NVIDIA's new Drive
  PX 2 AI platform for self-driving - Electrek}}.
\newblock
  \bibinfo{howpublished}{https://electrek.co/2016/10/21/all-new-teslas-are-equipped-with-nvidias-new-drive-px-2-ai-platform-for-self-driving}.
\newblock
\newblock
\shownote{[Online; accessed 9. Nov. 2020].}


\bibitem[\protect\citeauthoryear{Abuelsamid}{Abuelsamid}{2020}]%
        {Abuelsamid2020orin}
\bibfield{author}{\bibinfo{person}{Sam Abuelsamid}.}
  \bibinfo{year}{2020}\natexlab{}.
\newblock \showarticletitle{{Nvidia Cranks Up And Turns Down Its Drive AGX Orin
  Computers}}.
\newblock \bibinfo{journal}{\emph{Forbes}} (\bibinfo{date}{Jun}
  \bibinfo{year}{2020}).
\newblock
\urldef\tempurl%
\url{https://www.forbes.com/sites/samabuelsamid/2020/05/14/nvidia-cranks-up-and-turns-down-its-drive-agx-orin-computers}
\showURL{%
\tempurl}


\bibitem[\protect\citeauthoryear{Al~Faruque and Vatanparvar}{Al~Faruque and
  Vatanparvar}{2015}]%
        {al2015energy}
\bibfield{author}{\bibinfo{person}{Mohammad~Abdullah Al~Faruque} {and}
  \bibinfo{person}{Korosh Vatanparvar}.} \bibinfo{year}{2015}\natexlab{}.
\newblock \showarticletitle{Energy management-as-a-service over fog computing
  platform}.
\newblock \bibinfo{journal}{\emph{IEEE internet of things journal}}
  \bibinfo{volume}{3}, \bibinfo{number}{2} (\bibinfo{year}{2015}),
  \bibinfo{pages}{161--169}.
\newblock


\bibitem[\protect\citeauthoryear{ARM}{ARM}{[n.d.]}]%
        {arm-npu}
\bibfield{author}{\bibinfo{person}{ARM}.} \bibinfo{year}{[n.d.]}\natexlab{}.
\newblock \bibinfo{booktitle}{\emph{{arm npu Ethos-77}}}.
\newblock
\urldef\tempurl%
\url{https://www.arm.com/products/silicon-ip-cpu/ethos/ethos-n77}
\showURL{%
Retrieved April, 2021 from \tempurl}


\bibitem[\protect\citeauthoryear{Ba and Caruana}{Ba and Caruana}{2014}]%
        {KD1}
\bibfield{author}{\bibinfo{person}{Lei~Jimmy Ba} {and} \bibinfo{person}{Rich
  Caruana}.} \bibinfo{year}{2014}\natexlab{}.
\newblock \showarticletitle{Do Deep Nets Really Need to Be Deep?}. In
  \bibinfo{booktitle}{\emph{Proceedings of the 27th International Conference on
  Neural Information Processing Systems - Volume 2}} (Montreal, Canada)
  \emph{(\bibinfo{series}{NIPS'14})}. \bibinfo{publisher}{MIT Press},
  \bibinfo{address}{Cambridge, MA, USA}, \bibinfo{pages}{2654–2662}.
\newblock


\bibitem[\protect\citeauthoryear{Cho and Hariharan}{Cho and Hariharan}{2019}]%
        {KD4}
\bibfield{author}{\bibinfo{person}{Jang~Hyun Cho} {and}
  \bibinfo{person}{Bharath Hariharan}.} \bibinfo{year}{2019}\natexlab{}.
\newblock \showarticletitle{On the Efficacy of Knowledge Distillation}. In
  \bibinfo{booktitle}{\emph{Proceedings of the IEEE/CVF International
  Conference on Computer Vision (ICCV)}}.
\newblock


\bibitem[\protect\citeauthoryear{Codevilla, Lopez, Koltun, and
  Dosovitskiy}{Codevilla et~al\mbox{.}}{2018a}]%
        {codevilla2018offline}
\bibfield{author}{\bibinfo{person}{Felipe Codevilla},
  \bibinfo{person}{Antonio~M Lopez}, \bibinfo{person}{Vladlen Koltun}, {and}
  \bibinfo{person}{Alexey Dosovitskiy}.} \bibinfo{year}{2018}\natexlab{a}.
\newblock \showarticletitle{On offline evaluation of vision-based driving
  models}. In \bibinfo{booktitle}{\emph{Proceedings of the European Conference
  on Computer Vision (ECCV)}}. \bibinfo{pages}{236--251}.
\newblock


\bibitem[\protect\citeauthoryear{Codevilla, M{\"u}ller, L{\'o}pez, Koltun, and
  Dosovitskiy}{Codevilla et~al\mbox{.}}{2018b}]%
        {Codevilla2018}
\bibfield{author}{\bibinfo{person}{Felipe Codevilla}, \bibinfo{person}{Matthias
  M{\"u}ller}, \bibinfo{person}{Antonio L{\'o}pez}, \bibinfo{person}{Vladlen
  Koltun}, {and} \bibinfo{person}{Alexey Dosovitskiy}.}
  \bibinfo{year}{2018}\natexlab{b}.
\newblock \showarticletitle{End-to-end Driving via Conditional Imitation
  Learning}. In \bibinfo{booktitle}{\emph{International Conference on Robotics
  and Automation (ICRA)}}.
\newblock


\bibitem[\protect\citeauthoryear{Codevilla, Santana, L{\'o}pez, and
  Gaidon}{Codevilla et~al\mbox{.}}{2019}]%
        {codevilla2019exploring}
\bibfield{author}{\bibinfo{person}{Felipe Codevilla}, \bibinfo{person}{Eder
  Santana}, \bibinfo{person}{Antonio~M L{\'o}pez}, {and}
  \bibinfo{person}{Adrien Gaidon}.} \bibinfo{year}{2019}\natexlab{}.
\newblock \showarticletitle{Exploring the limitations of behavior cloning for
  autonomous driving}. In \bibinfo{booktitle}{\emph{Proceedings of the IEEE/CVF
  International Conference on Computer Vision}}. \bibinfo{pages}{9329--9338}.
\newblock


\bibitem[\protect\citeauthoryear{Cui, Zhong, Li, Chen, and Huang}{Cui
  et~al\mbox{.}}{2020}]%
        {cui2020offloading}
\bibfield{author}{\bibinfo{person}{Mingyue Cui}, \bibinfo{person}{Shipeng
  Zhong}, \bibinfo{person}{Boyang Li}, \bibinfo{person}{Xu Chen}, {and}
  \bibinfo{person}{Kai Huang}.} \bibinfo{year}{2020}\natexlab{}.
\newblock \showarticletitle{Offloading Autonomous Driving Services via Edge
  Computing}.
\newblock \bibinfo{journal}{\emph{IEEE Internet of Things Journal}}
  \bibinfo{volume}{7}, \bibinfo{number}{10} (\bibinfo{year}{2020}),
  \bibinfo{pages}{10535--10547}.
\newblock


\bibitem[\protect\citeauthoryear{Dosovitskiy, Ros, Codevilla, Lopez, and
  Koltun}{Dosovitskiy et~al\mbox{.}}{2017}]%
        {dosovitskiy2017carla}
\bibfield{author}{\bibinfo{person}{Alexey Dosovitskiy}, \bibinfo{person}{German
  Ros}, \bibinfo{person}{Felipe Codevilla}, \bibinfo{person}{Antonio Lopez},
  {and} \bibinfo{person}{Vladlen Koltun}.} \bibinfo{year}{2017}\natexlab{}.
\newblock \showarticletitle{CARLA: An open urban driving simulator}. In
  \bibinfo{booktitle}{\emph{Conference on robot learning}}. PMLR,
  \bibinfo{pages}{1--16}.
\newblock


\bibitem[\protect\citeauthoryear{Eichler}{Eichler}{2007}]%
        {eichler2007performance}
\bibfield{author}{\bibinfo{person}{Stephan Eichler}.}
  \bibinfo{year}{2007}\natexlab{}.
\newblock \showarticletitle{Performance evaluation of the IEEE 802.11 p WAVE
  communication standard}. In \bibinfo{booktitle}{\emph{2007 IEEE 66th
  Vehicular Technology Conference}}. IEEE, \bibinfo{pages}{2199--2203}.
\newblock


\bibitem[\protect\citeauthoryear{Feng, Liu, Wu, and Ji}{Feng
  et~al\mbox{.}}{2018}]%
        {feng2018mobile}
\bibfield{author}{\bibinfo{person}{Jingyun Feng}, \bibinfo{person}{Zhi Liu},
  \bibinfo{person}{Celimuge Wu}, {and} \bibinfo{person}{Yusheng Ji}.}
  \bibinfo{year}{2018}\natexlab{}.
\newblock \showarticletitle{Mobile edge computing for the internet of vehicles:
  Offloading framework and job scheduling}.
\newblock \bibinfo{journal}{\emph{IEEE vehicular technology magazine}}
  \bibinfo{volume}{14}, \bibinfo{number}{1} (\bibinfo{year}{2018}),
  \bibinfo{pages}{28--36}.
\newblock


\bibitem[\protect\citeauthoryear{González, Fang, Socarras, Serrat, Vázquez,
  Xu, and López}{González et~al\mbox{.}}{2016}]%
        {camera}
\bibfield{author}{\bibinfo{person}{Alejandro González},
  \bibinfo{person}{Zhijie Fang}, \bibinfo{person}{Yainuvis Socarras},
  \bibinfo{person}{Joan Serrat}, \bibinfo{person}{David Vázquez},
  \bibinfo{person}{Jiaolong Xu}, {and} \bibinfo{person}{Antonio~M. López}.}
  \bibinfo{year}{2016}\natexlab{}.
\newblock \showarticletitle{Pedestrian Detection at Day/Night Time with Visible
  and FIR Cameras: A Comparison}.
\newblock \bibinfo{journal}{\emph{Sensors}} \bibinfo{volume}{16},
  \bibinfo{number}{6} (\bibinfo{year}{2016}).
\newblock


\bibitem[\protect\citeauthoryear{He, Zhang, Ren, and Sun}{He
  et~al\mbox{.}}{2016}]%
        {resnet}
\bibfield{author}{\bibinfo{person}{Kaiming He}, \bibinfo{person}{Xiangyu
  Zhang}, \bibinfo{person}{Shaoqing Ren}, {and} \bibinfo{person}{Jian Sun}.}
  \bibinfo{year}{2016}\natexlab{}.
\newblock \showarticletitle{Deep Residual Learning for Image Recognition}. In
  \bibinfo{booktitle}{\emph{{CVPR} 2016}}. \bibinfo{publisher}{{IEEE} Computer
  Society}, \bibinfo{pages}{770--778}.
\newblock


\bibitem[\protect\citeauthoryear{Hecker, Dai, and Van~Gool}{Hecker
  et~al\mbox{.}}{2018}]%
        {hecker2018end}
\bibfield{author}{\bibinfo{person}{Simon Hecker}, \bibinfo{person}{Dengxin
  Dai}, {and} \bibinfo{person}{Luc Van~Gool}.} \bibinfo{year}{2018}\natexlab{}.
\newblock \showarticletitle{End-to-end learning of driving models with
  surround-view cameras and route planners}. In
  \bibinfo{booktitle}{\emph{Proceedings of the european conference on computer
  vision (eccv)}}. \bibinfo{pages}{435--453}.
\newblock


\bibitem[\protect\citeauthoryear{Hinton, Vinyals, and Dean}{Hinton
  et~al\mbox{.}}{2015}]%
        {KD2}
\bibfield{author}{\bibinfo{person}{Geoffrey Hinton}, \bibinfo{person}{Oriol
  Vinyals}, {and} \bibinfo{person}{Jeffrey Dean}.}
  \bibinfo{year}{2015}\natexlab{}.
\newblock \showarticletitle{Distilling the Knowledge in a Neural Network}. In
  \bibinfo{booktitle}{\emph{NIPS Deep Learning and Representation Learning
  Workshop}}.
\newblock
\urldef\tempurl%
\url{http://arxiv.org/abs/1503.02531}
\showURL{%
\tempurl}


\bibitem[\protect\citeauthoryear{Huang, Liu, and Weinberger}{Huang
  et~al\mbox{.}}{2016}]%
        {densenet}
\bibfield{author}{\bibinfo{person}{Gao Huang}, \bibinfo{person}{Zhuang Liu},
  {and} \bibinfo{person}{Kilian~Q. Weinberger}.}
  \bibinfo{year}{2016}\natexlab{}.
\newblock \showarticletitle{Densely Connected Convolutional Networks}.
\newblock \bibinfo{journal}{\emph{CoRR}}  \bibinfo{volume}{abs/1608.06993}
  (\bibinfo{year}{2016}).
\newblock


\bibitem[\protect\citeauthoryear{Huang et~al\mbox{.}}{Huang
  et~al\mbox{.}}{2012}]%
        {close}
\bibfield{author}{\bibinfo{person}{Junxian Huang} {et~al\mbox{.}}}
  \bibinfo{year}{2012}\natexlab{}.
\newblock \showarticletitle{A Close Examination of Performance and Power
  Characteristics of 4{G} LTE Networks}. In
  \bibinfo{booktitle}{\emph{Proceedings of the 10th International Conference on
  Mobile Systems, Applications, and Services}} \emph{(\bibinfo{series}{MobiSys
  '12})}. \bibinfo{pages}{225–238}.
\newblock


\bibitem[\protect\citeauthoryear{Hyatt}{Hyatt}{2019}]%
        {baiduapollo}
\bibfield{author}{\bibinfo{person}{Kyle Hyatt}.}
  \bibinfo{year}{2019}\natexlab{}.
\newblock \showarticletitle{{Baidu unveils its camera-based Apollo Lite
  self-driving suite}}.
\newblock \bibinfo{journal}{\emph{Roadshow}} (\bibinfo{date}{Jun}
  \bibinfo{year}{2019}).
\newblock
\urldef\tempurl%
\url{https://www.cnet.com/roadshow/news/baidu-apollo-lite-camera-based-self-driving}
\showURL{%
\tempurl}


\bibitem[\protect\citeauthoryear{Hyatt}{Hyatt}{2021}]%
        {argoai}
\bibfield{author}{\bibinfo{person}{Kyle Hyatt}.}
  \bibinfo{year}{2021}\natexlab{}.
\newblock \showarticletitle{{Argo gives its self-driving vehicle hardware a big
  upgrade}}.
\newblock \bibinfo{journal}{\emph{Roadshow}} (\bibinfo{date}{Jan}
  \bibinfo{year}{2021}).
\newblock
\urldef\tempurl%
\url{https://www.cnet.com/roadshow/news/argo-self-driving-car-hardware-upgrade}
\showURL{%
\tempurl}


\bibitem[\protect\citeauthoryear{Imoize, Orolu, and Atayero}{Imoize
  et~al\mbox{.}}{2020}]%
        {imoize2020analysis}
\bibfield{author}{\bibinfo{person}{Agbotiname~Lucky Imoize},
  \bibinfo{person}{Kehinde Orolu}, {and} \bibinfo{person}{Aderemi Aaron-Anthony
  Atayero}.} \bibinfo{year}{2020}\natexlab{}.
\newblock \showarticletitle{Analysis of key performance indicators of a 4G LTE
  network based on experimental data obtained from a densely populated smart
  city}.
\newblock \bibinfo{journal}{\emph{Data in brief}}  \bibinfo{volume}{29}
  (\bibinfo{year}{2020}), \bibinfo{pages}{105304}.
\newblock


\bibitem[\protect\citeauthoryear{Kang, Hauswald, Gao, Rovinski, Mudge, Mars,
  and Tang}{Kang et~al\mbox{.}}{2017}]%
        {neurosurgeon}
\bibfield{author}{\bibinfo{person}{Yiping Kang}, \bibinfo{person}{Johann
  Hauswald}, \bibinfo{person}{Cao Gao}, \bibinfo{person}{Austin Rovinski},
  \bibinfo{person}{Trevor Mudge}, \bibinfo{person}{Jason Mars}, {and}
  \bibinfo{person}{Lingjia Tang}.} \bibinfo{year}{2017}\natexlab{}.
\newblock \showarticletitle{Neurosurgeon: Collaborative Intelligence Between
  the Cloud and Mobile Edge}. In \bibinfo{booktitle}{\emph{Proceedings of the
  Twenty-Second International Conference on Architectural Support for
  Programming Languages and Operating Systems}} \emph{(\bibinfo{series}{ASPLOS
  '17})}. \bibinfo{pages}{615–629}.
\newblock


\bibitem[\protect\citeauthoryear{{Khayyat}, {Elgendy}, {Muthanna},
  {Alshahrani}, {Alharbi}, and {Koucheryavy}}{{Khayyat} et~al\mbox{.}}{2020}]%
        {RL1}
\bibfield{author}{\bibinfo{person}{M. {Khayyat}}, \bibinfo{person}{I.~A.
  {Elgendy}}, \bibinfo{person}{A. {Muthanna}}, \bibinfo{person}{A.~S.
  {Alshahrani}}, \bibinfo{person}{S. {Alharbi}}, {and} \bibinfo{person}{A.
  {Koucheryavy}}.} \bibinfo{year}{2020}\natexlab{}.
\newblock \showarticletitle{Advanced Deep Learning-Based Computational
  Offloading for Multilevel Vehicular Edge-Cloud Computing Networks}.
\newblock \bibinfo{journal}{\emph{IEEE Access}}  \bibinfo{volume}{8}
  (\bibinfo{year}{2020}), \bibinfo{pages}{137052--137062}.
\newblock


\bibitem[\protect\citeauthoryear{Kim, Son, Song, and Kim}{Kim
  et~al\mbox{.}}{2018}]%
        {radar}
\bibfield{author}{\bibinfo{person}{Young-Duk Kim}, \bibinfo{person}{Guk-Jin
  Son}, \bibinfo{person}{Chan-Ho Song}, {and} \bibinfo{person}{Hee-Kang Kim}.}
  \bibinfo{year}{2018}\natexlab{}.
\newblock \showarticletitle{On the Deployment and Noise Filtering of Vehicular
  Radar Application for Detection Enhancement in Roads and Tunnels}.
\newblock \bibinfo{journal}{\emph{Sensors}} \bibinfo{volume}{18},
  \bibinfo{number}{3} (\bibinfo{year}{2018}).
\newblock


\bibitem[\protect\citeauthoryear{Kong}{Kong}{2020}]%
        {kong2020computation}
\bibfield{author}{\bibinfo{person}{Peng-Yong Kong}.}
  \bibinfo{year}{2020}\natexlab{}.
\newblock \showarticletitle{Computation and sensor offloading for cloud-based
  infrastructure-assisted autonomous vehicles}.
\newblock \bibinfo{journal}{\emph{IEEE Systems Journal}} \bibinfo{volume}{14},
  \bibinfo{number}{3} (\bibinfo{year}{2020}), \bibinfo{pages}{3360--3370}.
\newblock


\bibitem[\protect\citeauthoryear{Li, Yang, Xie, Li, and Xu}{Li
  et~al\mbox{.}}{2018}]%
        {lidar}
\bibfield{author}{\bibinfo{person}{Xiaolu Li}, \bibinfo{person}{Bingwei Yang},
  \bibinfo{person}{Xinhao Xie}, \bibinfo{person}{Duan Li}, {and}
  \bibinfo{person}{Lijun Xu}.} \bibinfo{year}{2018}\natexlab{}.
\newblock \showarticletitle{Influence of Waveform Characteristics on LiDAR
  Ranging Accuracy and Precision}.
\newblock \bibinfo{journal}{\emph{Sensors}} \bibinfo{volume}{18},
  \bibinfo{number}{4} (\bibinfo{year}{2018}).
\newblock


\bibitem[\protect\citeauthoryear{Lin, Zhang, Hsu, Skach, Haque, Tang, and
  Mars}{Lin et~al\mbox{.}}{2018}]%
        {lin2018architectural}
\bibfield{author}{\bibinfo{person}{Shih-Chieh Lin}, \bibinfo{person}{Yunqi
  Zhang}, \bibinfo{person}{Chang-Hong Hsu}, \bibinfo{person}{Matt Skach},
  \bibinfo{person}{Md~E Haque}, \bibinfo{person}{Lingjia Tang}, {and}
  \bibinfo{person}{Jason Mars}.} \bibinfo{year}{2018}\natexlab{}.
\newblock \showarticletitle{The architectural implications of autonomous
  driving: Constraints and acceleration}. In
  \bibinfo{booktitle}{\emph{Proceedings of the Twenty-Third International
  Conference on Architectural Support for Programming Languages and Operating
  Systems}}. \bibinfo{pages}{751--766}.
\newblock


\bibitem[\protect\citeauthoryear{Matsubara, Baidya, Callegaro, Levorato, and
  Singh}{Matsubara et~al\mbox{.}}{2019}]%
        {marco2}
\bibfield{author}{\bibinfo{person}{Yoshitomo Matsubara}, \bibinfo{person}{Sabur
  Baidya}, \bibinfo{person}{Davide Callegaro}, \bibinfo{person}{Marco
  Levorato}, {and} \bibinfo{person}{Sameer Singh}.}
  \bibinfo{year}{2019}\natexlab{}.
\newblock \showarticletitle{Distilled Split Deep Neural Networks for
  Edge-Assisted Real-Time Systems}. In \bibinfo{booktitle}{\emph{Proceedings of
  the 2019 Workshop on Hot Topics in Video Analytics and Intelligent Edges}}
  (Los Cabos, Mexico) \emph{(\bibinfo{series}{HotEdgeVideo'19})}.
  \bibinfo{publisher}{Association for Computing Machinery},
  \bibinfo{address}{New York, NY, USA}, \bibinfo{pages}{21–26}.
\newblock
\showISBNx{9781450369282}


\bibitem[\protect\citeauthoryear{{Matsubara}, {Callegaro}, {Baidya},
  {Levorato}, and {Singh}}{{Matsubara} et~al\mbox{.}}{2020}]%
        {marco1}
\bibfield{author}{\bibinfo{person}{Y. {Matsubara}}, \bibinfo{person}{D.
  {Callegaro}}, \bibinfo{person}{S. {Baidya}}, \bibinfo{person}{M. {Levorato}},
  {and} \bibinfo{person}{S. {Singh}}.} \bibinfo{year}{2020}\natexlab{}.
\newblock \showarticletitle{Head Network Distillation: Splitting Distilled Deep
  Neural Networks for Resource-Constrained Edge Computing Systems}.
\newblock \bibinfo{journal}{\emph{IEEE Access}}  \bibinfo{volume}{8}
  (\bibinfo{year}{2020}), \bibinfo{pages}{212177--212193}.
\newblock


\bibitem[\protect\citeauthoryear{Matsubara and Levorato}{Matsubara and
  Levorato}{2020}]%
        {marco3}
\bibfield{author}{\bibinfo{person}{Yoshitomo Matsubara} {and}
  \bibinfo{person}{Marco Levorato}.} \bibinfo{year}{2020}\natexlab{}.
\newblock \bibinfo{title}{Neural Compression and Filtering for Edge-assisted
  Real-time Object Detection in Challenged Networks}.
\newblock
\newblock
\showeprint[arxiv]{2007.15818}~[cs.CV]


\bibitem[\protect\citeauthoryear{Odema, Rashid, Demirel, and Faruque}{Odema
  et~al\mbox{.}}{2021}]%
        {odema2021lens}
\bibfield{author}{\bibinfo{person}{Mohanad Odema}, \bibinfo{person}{Nafiul
  Rashid}, \bibinfo{person}{Berken~Utku Demirel}, {and}
  \bibinfo{person}{Mohammad Abdullah~Al Faruque}.}
  \bibinfo{year}{2021}\natexlab{}.
\newblock \showarticletitle{LENS: Layer Distribution Enabled Neural
  Architecture Search in Edge-Cloud Hierarchies}. In
  \bibinfo{booktitle}{\emph{2021 58th ACM/IEEE Design Automation Conference
  (DAC)}}.
\newblock


\bibitem[\protect\citeauthoryear{Oh}{Oh}{2017}]%
        {Oh2017pegasus}
\bibfield{author}{\bibinfo{person}{Nate Oh}.} \bibinfo{year}{2017}\natexlab{}.
\newblock \showarticletitle{{NVIDIA Announces Drive PX Pegasus at GTC Europe
  2017: Level 5 Self-Driving Hardware, Feat. Post-Volta GPUs}}.
\newblock \bibinfo{journal}{\emph{AnandTech}} (\bibinfo{date}{Oct}
  \bibinfo{year}{2017}).
\newblock
\urldef\tempurl%
\url{https://www.anandtech.com/show/11913/nvidia-announces-drive-px-pegasus-at-gtc-europe-2017-feat-nextgen-gpus}
\showURL{%
\tempurl}


\bibitem[\protect\citeauthoryear{Papathanassiou and Khoryaev}{Papathanassiou
  and Khoryaev}{2017}]%
        {papathanassiou2017cellular}
\bibfield{author}{\bibinfo{person}{Apostolos Papathanassiou} {and}
  \bibinfo{person}{Alexey Khoryaev}.} \bibinfo{year}{2017}\natexlab{}.
\newblock \showarticletitle{Cellular V2X as the essential enabler of superior
  global connected transportation services}.
\newblock \bibinfo{journal}{\emph{IEEE 5G Tech Focus}} \bibinfo{volume}{1},
  \bibinfo{number}{2} (\bibinfo{year}{2017}), \bibinfo{pages}{1--2}.
\newblock


\bibitem[\protect\citeauthoryear{{Samal}, {Wolf}, and {Mukhopadhyay}}{{Samal}
  et~al\mbox{.}}{2020}]%
        {pruning}
\bibfield{author}{\bibinfo{person}{K. {Samal}}, \bibinfo{person}{M. {Wolf}},
  {and} \bibinfo{person}{S. {Mukhopadhyay}}.} \bibinfo{year}{2020}\natexlab{}.
\newblock \showarticletitle{Attention-Based Activation Pruning to Reduce Data
  Movement in Real-Time AI: A Case-Study on Local Motion Planning in Autonomous
  Vehicles}.
\newblock \bibinfo{journal}{\emph{IEEE Journal on Emerging and Selected Topics
  in Circuits and Systems}} \bibinfo{volume}{10}, \bibinfo{number}{3}
  (\bibinfo{year}{2020}), \bibinfo{pages}{306--319}.
\newblock


\bibitem[\protect\citeauthoryear{Sasaki, Suzuki, Makido, and Nakao}{Sasaki
  et~al\mbox{.}}{2016}]%
        {sasaki2016vehicle}
\bibfield{author}{\bibinfo{person}{Kengo Sasaki}, \bibinfo{person}{Naoya
  Suzuki}, \bibinfo{person}{Satoshi Makido}, {and} \bibinfo{person}{Akihiro
  Nakao}.} \bibinfo{year}{2016}\natexlab{}.
\newblock \showarticletitle{Vehicle control system coordinated between cloud
  and mobile edge computing}. In \bibinfo{booktitle}{\emph{2016 55th Annual
  Conference of the Society of Instrument and Control Engineers of Japan
  (SICE)}}. IEEE, \bibinfo{pages}{1122--1127}.
\newblock


\bibitem[\protect\citeauthoryear{Tampuu, Matiisen, Semikin, Fishman, and
  Muhammad}{Tampuu et~al\mbox{.}}{2020}]%
        {tampuu2020survey}
\bibfield{author}{\bibinfo{person}{Ardi Tampuu}, \bibinfo{person}{Tambet
  Matiisen}, \bibinfo{person}{Maksym Semikin}, \bibinfo{person}{Dmytro
  Fishman}, {and} \bibinfo{person}{Naveed Muhammad}.}
  \bibinfo{year}{2020}\natexlab{}.
\newblock \showarticletitle{A survey of end-to-end driving: Architectures and
  training methods}.
\newblock \bibinfo{journal}{\emph{IEEE Transactions on Neural Networks and
  Learning Systems}} (\bibinfo{year}{2020}).
\newblock


\bibitem[\protect\citeauthoryear{Urban, Geras, Kahou, Aslan, Wang, Caruana,
  Mohamed, Philipose, and Richardson}{Urban et~al\mbox{.}}{2017}]%
        {KD3}
\bibfield{author}{\bibinfo{person}{Gregor Urban}, \bibinfo{person}{Krzysztof~J.
  Geras}, \bibinfo{person}{Samira~Ebrahimi Kahou}, \bibinfo{person}{Ozlem
  Aslan}, \bibinfo{person}{Shengjie Wang}, \bibinfo{person}{Rich Caruana},
  \bibinfo{person}{Abdelrahman Mohamed}, \bibinfo{person}{Matthai Philipose},
  {and} \bibinfo{person}{Matt Richardson}.} \bibinfo{year}{2017}\natexlab{}.
\newblock \bibinfo{title}{Do Deep Convolutional Nets Really Need to be Deep and
  Convolutional?}
\newblock
\newblock
\showeprint[arxiv]{1603.05691}~[stat.ML]


\bibitem[\protect\citeauthoryear{Vatanparvar and Faruque}{Vatanparvar and
  Faruque}{2018}]%
        {vatanparvar2018design}
\bibfield{author}{\bibinfo{person}{Korosh Vatanparvar} {and}
  \bibinfo{person}{Mohammad Abdullah~Al Faruque}.}
  \bibinfo{year}{2018}\natexlab{}.
\newblock \showarticletitle{Design and analysis of battery-aware automotive
  climate control for electric vehicles}.
\newblock \bibinfo{journal}{\emph{ACM Transactions on Embedded Computing
  Systems (TECS)}} \bibinfo{volume}{17}, \bibinfo{number}{4}
  (\bibinfo{year}{2018}), \bibinfo{pages}{1--22}.
\newblock


\bibitem[\protect\citeauthoryear{Xie and Kim}{Xie and Kim}{2019}]%
        {jpeg}
\bibfield{author}{\bibinfo{person}{Xiufeng Xie} {and} \bibinfo{person}{Kyu-Han
  Kim}.} \bibinfo{year}{2019}\natexlab{}.
\newblock \showarticletitle{Source Compression with Bounded DNN Perception Loss
  for IoT Edge Computer Vision}. In \bibinfo{booktitle}{\emph{The 25th Annual
  International Conference on Mobile Computing and Networking}} (Los Cabos,
  Mexico) \emph{(\bibinfo{series}{MobiCom '19})}.
  \bibinfo{publisher}{Association for Computing Machinery},
  \bibinfo{address}{New York, NY, USA}, Article \bibinfo{articleno}{47},
  \bibinfo{numpages}{16}~pages.
\newblock


\bibitem[\protect\citeauthoryear{Zhang, Mao, Leng, Maharjan, and Zhang}{Zhang
  et~al\mbox{.}}{2017}]%
        {zhang2017optimal}
\bibfield{author}{\bibinfo{person}{Ke Zhang}, \bibinfo{person}{Yuming Mao},
  \bibinfo{person}{Supeng Leng}, \bibinfo{person}{Sabita Maharjan}, {and}
  \bibinfo{person}{Yan Zhang}.} \bibinfo{year}{2017}\natexlab{}.
\newblock \showarticletitle{Optimal delay constrained offloading for vehicular
  edge computing networks}. In \bibinfo{booktitle}{\emph{2017 IEEE
  International Conference on Communications (ICC)}}. IEEE,
  \bibinfo{pages}{1--6}.
\newblock


\bibitem[\protect\citeauthoryear{Zhang, Mao, Leng, Zhao, Li, Peng, Pan,
  Maharjan, and Zhang}{Zhang et~al\mbox{.}}{2016}]%
        {zhang2016energy}
\bibfield{author}{\bibinfo{person}{Ke Zhang}, \bibinfo{person}{Yuming Mao},
  \bibinfo{person}{Supeng Leng}, \bibinfo{person}{Quanxin Zhao},
  \bibinfo{person}{Longjiang Li}, \bibinfo{person}{Xin Peng},
  \bibinfo{person}{Li Pan}, \bibinfo{person}{Sabita Maharjan}, {and}
  \bibinfo{person}{Yan Zhang}.} \bibinfo{year}{2016}\natexlab{}.
\newblock \showarticletitle{Energy-efficient offloading for mobile edge
  computing in 5G heterogeneous networks}.
\newblock \bibinfo{journal}{\emph{IEEE access}}  \bibinfo{volume}{4}
  (\bibinfo{year}{2016}), \bibinfo{pages}{5896--5907}.
\newblock


\end{thebibliography}


\end{document}